\documentclass[sts]{imsart}

\RequirePackage{amsthm,amsmath,amsfonts,amssymb, MnSymbol}
\RequirePackage[numbers]{natbib}
\RequirePackage[colorlinks,citecolor=blue,urlcolor=blue]{hyperref}
\RequirePackage{graphicx}

\startlocaldefs
\theoremstyle{plain}

\theoremstyle{remark}

\endlocaldefs

\begin{document}

\begin{frontmatter}
\title{A Cheat Sheet for Bayesian Prediction}

\begin{aug}
\author[A]{\fnms{Bertrand} \snm{Clarke}\ead[label=e1]{bclarke3@unl.edu}}
\and
\author[B]{\fnms{Yuling} \snm{Yao}\ead[label=e2]{yyao@flatironinstitute.org}}

\address[A]{Bertrand Clarke is Professor, Department of Statstics,
University of Nebraska-Lincoln,  USA \printead{e1}.}

\address[B]{Yuling Yao is a post-doctoral scholar at the Flatiron
Institute, New York. \printead{e2}.}

\end{aug}

\begin{abstract}

This paper reviews the growing field of Bayesian prediction.
Bayes point and interval prediction are defined and exemplified and
situated in statistical prediction more generally.
Then, four general approaches to Bayes prediction are defined
 and we turn to predictor selection.  This can be done
predictively or non-predictively and predictors can be based on
single models or multiple models.  We call these latter cases unitary
predictors and model average predictors, respectively.
Then we turn to the most recent aspect of prediction to emerge, namely
 prediction in the context of large observational data sets and discuss 
three further classes of techniques.
We conclude with a summary and statement of several current open problems.

\end{abstract}

\end{frontmatter}

\section{Formulating the Predictive Problem}
\label{Intro}

The central goal of Bayesian prediction is to give useful expressions for
\begin{eqnarray}
P( Y_{n+1} \in A \vert Y^n = y^n).
\label{cdldtn}
\end{eqnarray}
In \eqref{cdldtn},  $Y^n = (Y_1, \ldots , Y_n)^T$ is a random vector 
with outcomes $y^n = (y_1, \ldots, y_n)^T$,  $Y_{n+1}$ is the random variable we are trying to predict 
and $A$ is a measureable set.  One reason this 
is Bayesian is that the operation
indicated by `$ \vert$' is conditioning meaning that the $Y_i$'s are dependent 
and the entire prediction problem is defined on one measureable space i.e., the `containment principle'
of Bayesian statistics is satisfied.  Another reason this is Bayesian is that we seek
an entire probability measure; a point predictor is not enough.   Although point predictors may be
Bayesian, too, it is in a different sense.   Finally, to be Bayesian, predictors of the form \eqref{cdldtn} must be
Bayes optimal in some sense, see  Sec.  \ref{paradigms} for a discussion of this
point. An important secondary goal of Bayesian prediction is to assess the comparative 
performance of expressions
of the form \eqref{cdldtn} through their theoretical properties and their numerical
performance.

The most basic instantiation of \eqref{cdldtn} is the following.  Suppose $Y_i$ for $i=1, \ldots, n+1$
are independently and identically distributed (IID) with density $p(\cdot \vert \theta)$
(with respect to Lebesgue measure) given the finite dimensional parameter
$\theta = (\theta_1, \ldots , \theta_p) \in \Omega$ where $\Omega \subset \mathbb{R}^p$ is open.
Assume $\theta \sim W$ where the prior probability $W$ has density $w$
with respect to Lebesgue measure.  By Bayes rule,  the form of \eqref{cdldtn} in this case is
\begin{eqnarray}
P( Y_{n+1} \in A \vert Y^n = y^n) &=& M_n (A\vert y^n) 
\nonumber \\
&=& \int_A m_n(y_{n+1} \vert y^n)  {\rm d} y_{n+1} 
\label{postpred1}
\end{eqnarray}
in which the density $m_n(y_{n+1} \vert y^n)$ of the conditional mixture probability $M_n(\cdot \vert y^n)$ is
\begin{eqnarray}
m_n(y_{n+1} \vert y^n) =   \int_\Omega p(y_{n+1} \vert \theta)w(\theta \vert y^n) {\rm d} \theta 
\label{postpred2}
\end{eqnarray}
and $w(\theta \vert y^n) \propto w(\theta) p(y^n \vert \theta)$ is the posterior density.
Now we get a 100$(1-\alpha)$ predictive interval (PI)
\begin{eqnarray}
M_{n+1}( Y_{n+1} \in [t_\alpha/2 , t_{1 - \alpha/2}] \vert y^n) = 1 - \alpha
\label{1stPI}
\end{eqnarray}
by taking $t_{\alpha/2}$ and $t_{1-\alpha/2}$ to be the 100$(\alpha/2)$\% and 100 ($1-\alpha/2$) percentiles of
$M_{n+1}( \cdot \vert y^n)$.

If $p_\theta$ is the true density and we have data $y^n$, the conditional mixture density is 
Bayes optimal for prediction
in Kullback-Leibler distance.   Following \cite{Aitchison:1975}, it is easy to 
see this:  Let $D(P \| Q) = \int p(x) \ln (p(x)/q(x)) {\rm d} \mu(x)$ where
$p$ and $q$ are the densities of probability measures $P$ and $Q$ with respect to $\mu$, here
taken to be Lebesgue measure.   Then writing $P_\theta$ for the probability associated to $p_\theta$
and for $q(\cdot)$ any density with respect to Lebesgue measure with probability $Q$
for which the terms are finite, we have
\begin{eqnarray}
 &&\int_\Omega  D(P_\theta \| Q )w(\theta \vert y^n) {\rm d} \theta
\nonumber \\
&=&
\int_\Omega  D(P_\theta \| M(\cdot \vert y^n ))w(\theta \vert y^n) {\rm d} \theta
\nonumber \\
&+&
\int_\Omega \int _{\mathbb{R}} p_\theta(y_{n+1} ) \ln \frac{m(y_{n+1} \vert y^n)}{q(y_{n+1})} w(\theta \vert y^n) {\rm d} \theta
\nonumber \\
&=& I(\Theta; Y_{n+1} \vert Y^n = y^n) 
\nonumber \\
&& + \int m(y_{n+1} \vert y^n) \ln \frac{m(y_{n+1} \vert y^n) }{q(y_{n+1})} {\rm d} y_{n+1}.
\label{KLopt}
\end{eqnarray}
In \eqref{KLopt}, the first term on the right is the Shannon (conditional) mutual information and
is non-negative.
The second term is also non-negative.  So, the left hand side is minimized when
$Q$ is obtained from $q(\cdot) = m(\cdot \vert y^n)$.  Thus, the conditional mixture is the
optimal distribution from which to derive $(1-\alpha)$ PI's such as \eqref{1stPI}.

Let's look at an entertaining example from \cite{Galton:1907}.  Originally titled
{\it Vox Populi} to suggest the pooling of opinions, the principle is now often 
called the `Wisdom of Crowds'.  See \cite{Wallis:2014} for a predictive discussion
of Galton's Ox problem.   We will see how expressions such as \eqref{postpred2}
can be combined into useful expressions for a point predictor version of \eqref{cdldtn}.
Suppose $n=0$ and $y^0 = \{ \varnothing  \}$.  Then there is no past data and we are in 
round one of a sequential prediction problem,
i.e., trying to predict $Y_1$.  We are treating the dressed weight of the ox as a random
variable, implicitly assuing it was drawn from a population of oxen.

The `experiment' was in 1907.  
Approximately 800 men guessed the dressed weight of an ox, $Y$.
Each man had to pay a small fee to issue his guess $\hat{Y}_i= \hat{y}_i$
as to the weight and whoever's guess was closest would win a prize.  Without strain we can assume the men
made their guesses independently -- who would want to share information at
the cost of possibly not winning the prize?  We cannot assume that all the
people issuing guesses were equally capable of issuing {\it good} guesses, but
we can assume each did his best.   It's possible someone could have
purposefully issued misleading information 
so as to hamper others, but for simplicity we'll assume no one did this.
In fact, 13 of the guesses were disqualified for various reasons leaving 787 weights 
(in pounds).  Thus, the data consisted of
787 guesses -- predictions, $\hat{y}_i$  -- and one measured weight to be predicted
i.e.,  the observation $y_1$.

Let us regard $\hat{y}_i$ as having come from the $i$-th man's posterior predictive, say
\begin{eqnarray}
\hat{y}_i &=& \int     y_i \int_{\Omega_i} p_i(y_i \vert \theta_i) w_i(\theta_i \vert y^0) {\rm d} y_i
\nonumber \\
&=&  \int     y_i \int_{\Omega_i} p_i(y_i \vert \theta_i) w_i(\theta_i){\rm d} \theta_i  {\rm d} y_i
\nonumber \\
&=& \int y_i m_i(y_i) {\rm d} y_i
\label{ithman}
\end{eqnarray}
where \eqref{ithman} is the integral of \eqref{postpred2} and
each quantity (prior, likelihood,parameter, and parameter space) is unique to man $i$.
We may also write $p_i(y_i \vert \theta_i, x_i)$ if we believe that man $i$ has access to
explanatory variables $x_i$.
In effect we are assuming each man had a model $p_i ( \cdot \vert \theta_i)$ and prior $w_i(\theta_i)$
in his head and that he used squared error loss (see Subsec. \ref{decisiontheory})
to justify taking the mean over $y_i$ in \eqref{ithman}
to get his prediction.  The prior and likelihood are inaccessible to us making this an
${\cal{M}}$-complete problem, see Subec. \ref{probclasses}.

To see how variable the implied $p_i$'s were, \cite{Galton:1907} made a graph of the $\hat{y}_i$'s.
He found that the middle 50\% were distributed, roughly as a truncated normal, the upper and lower 25\% 
gave asymmetry with the lower values stretched out further than the upper values.
While \cite{Galton:1907} argued that the median of the $\hat{y}_i$'s was a good predictor of $Y_1 = y_1$
perhaps thinking of abslute error on account of the asymmetry,
\cite{Wallis:2014} noted that in this particular instance, the mean of the $\hat{y}_i$'s was actually closer to
$y_i$ than the median.  That is, \cite{Wallis:2014} argued that
\begin{eqnarray}
\frac{1}{787} \sum_{i=1}^{787} \hat{y}_i =  \sum_{i=1}^{787} W(i \vert y^0) \int y_i m_i(y_i) {\rm d} y_i
\label{meanofpreds}
\end{eqnarray}
where $W$ is the across-model prior weighting the predictions.  Here, $W$ is taken to be uniform since
there is no reason to weight one prediction more than another and there is no data to condition on.
Expression \eqref{meanofpreds} is an example of a model average predictor, see  \eqref{postpred3}
in Subsec. \ref{Bayesrule} where the predictions are averaged using the \footnotemark posterior model weights
(even though here the posterior is trivial). 

\footnotetext{A median can also be used to give a model average, see Sec. \ref{Example1}.}

If we insisted on a PI, we could identify $w_i$, $\theta_i$,  and $p_i$ for man $i$ and then
form an $i$-th PI as in \eqref{1stPI}.  If we insisted on a PI that combined
the 787 PI's, this, too, is possible,  see Sec. \ref{ModelAverages}.
That combining individual predictions generally gives a better prediction has been known for a long time,
see \cite{Bates:Granger:1969} who perhaps did the most to popularise the idea.  See
\cite{Clemen:1989} who gives an annotated bibliography supporting this idea; some references
are as early as 1926.  Our treatment differs from \cite{Tian:etal:2022} because ours is purely
Bayesian and 
conceptual.

The remainder of this paper is structured as follows.  Sec. \ref{background}
provides some of the more abstract concepts undergiurding the predcitive view.
Sec.  \ref{paradigms} outlines the four main Bayesian philosophies of prediction,
and the techniques they motivate,
and briefly mentions four more.
Sec. \ref{Predsel} looks at the selection of individual predictors, here called unitary.
Sec. \ref{ModelAverages} looks at model averaging predictors i.e., not unitary.  
Sec.  \ref{streaming} looks at prediction with streaming data.   Sec. \ref{conclusion} concludes with
a discussion of where Bayes prediction is and appears to be headeed as well as
serving to introduce the papers in this issue.

\section{Conceptual Background}
\label{background}

Given the specific cases in Sec. \ref{Intro}, it is worth outlining
the generalities of the predictive framework before discussing the 
main philosophies, and the techniques they suggest.  There are two
key concepts -- the relationship between predictions and data and
the relation between predictors and classes of predictors.

\subsection{Validation and Calibration}
\label{valcal}

It is  not enough to have \eqref{cdldtn}.  The predictor must validate; if it doesn't, it is discredited.
Validation means that the predictions the predictor gives are accurate to within the uncertianty
of the predictor.
In particular, If the predictor is based on a model and fails to validate the model is discredited.
So, as a generality, in practice, prediction is less forgiving than modeling, testing, and estimation because
poor prediction is usually more immediate and obvious.   Box's dictum
all models are wrong but some are useful has no good analog in prediction, conceptually, because the
range of wrong models is so vast that a practitioner is likely to have one that will easily
be discredited.
Indeed, prediction is so fundamental that a model's worth is primarily a function of how
successful the predictors it generates are.    The predictive counterpart to Box's dictum
might be:  All predictors are guilty of
bias and incorrect assessment of variance until proven innocent and
the degree of predictive success a method has
constrains the reliability of any interpretation that rests
on it.  

In the Bayesian context, validation is closely related to calibration.  It is well-known that within
the Bayesian paradigm, bias cannot be detected;  see \cite{Berk:1966} who showed that 
a posterior distribution will converge but not necessarily to the data generator.  
Moreover,
calibration is an extra procedure that
must be successfully completed to ensure good inference, see \cite{Dawid:1982}.   In practice, this
is hard especially in cases where the internal components of the data generator (DG) are not 
stable enough to allow accurate modeling.   Another difficulty is that even if calibration can be
done accurately in a modeling sense, often sufficent data required for calibration are unavailable.
One of the benefits of the purely predictive approach is that predictors can be reformulated
in view of accumulated data.  That is, by discrediting a predictor, validation failure automatically 
forces reselection of the predictor, perhaps by examination of the pattern of errors.

Forming a predictor may require forming an estimator.  However, the usefulness
of the estimator is only in terms of how well it facilitates good prediction; its 
worth purely an as estimator is conceptually unimportant.  By contrast, in the estimation view, the quantity being
estimated is assumed well-defined, physically meaningful, and important independent
of its use in prediction.  In fact, estimators are rarely predictively validated and hence
their validity typically rests on physical modeling assumptions.  

This is rarely
enough to give good prediction.  Indeed, with complex DG's detailed
modeling is effectively impossible so validated predictors and predictions are
most of what the analyst has to make inferences.  Other factors might include stability,
complexity, risk, etc.  The net effect of this is that the predictive view is
more foundational than the estimation view since good prediction does not
directly require good modeling.  That is, validated predictors are a better
summarization of the data than a convincing model that has not been
predictively validated.

One counterargument to this is that models can be validated by goodness of fit
testing.   Conditional on a model with a finite dimension parameter being believed true,
for instance,
this argument is supportive of the estimation view.  However, once one admits that the
model is false -- per Box's admission --  the class of all possible and `reasonable'
models for a given sample size is usually very large.  Indeed,  the space of non-rejectable models is 
often so large that the range of predictions it can give is too large to be useful.
Absent vaalidation,  a model is only a complicated summary statistic for a data set not unlike
having a collection of, say, sample quantiles.

A related problem is that, fundamentally, goodness of fit of a model is inadequate
as a criterion.  Indeed, a model is only `scientific' if it can make predictions
about observables that can be measured and so that bad enough
prediction can be used to discredit the model.   Thus, sufficent goodness
of fit is evidence in favor of a model but by itself does not rise to the level
of validation.   Here, when we use the term model we assume
that it leads to a unique and tractable predictor. 

Obviously, \eqref{cdldtn} must be generalized to include explanatory variables.
Assume we have data
${\cal{D}} = {\cal{D}}_n =\{(x_i, y_i) \vert  i=1, \ldots , n\}$ where each $x_i$ is an
instance of an explanatory vector $(x_{1i}, \ldots ,x_{d i})^T$.
Now a PI for the $n+1$ outcome $Y_{n+1}$ at $x_{n+1}$ is 
\begin{eqnarray}
P( Y_{n+1}(x_{n+1}) \in [a, b] \mid {\cal{D}}) ) \geq 1 - \alpha ,
\label{PIdef}
\end{eqnarray}
where $a = a({\cal{D}})$ and $b = b({\cal{D}})$.   This restricts $A$ in \eqref{cdldtn} 
to intervals, but that is usually sufficient and \eqref{PIdef} can easily be modified to accommodate
more complex sets.  We also often write $\hat{Y}_{n+1}(x)$ to mean a
point predictor for $Y_{n+1}(x)$ where it is understood that $\hat{Y}_{n+1}(x)$ is also
a function of ${\cal{D}}$ and we want
$\hat{Y}_{n+1}(x_{n+1})$ to be close to 
$Y_{n+1}(x_{n+1})$ in some formal sense, e.g., 
$P( \vert \hat{Y}_{n+1}(x_{n+1}) - Y_{n+1}(x_{n+1}) \vert \leq K_1\sigma^2(1 + K_2/n) \vert {\cal{D}} )$ is small for
some $K_1, K_2>0$ under the true probability $P$,
where ${\sf Var}(Y_{n+1}(x_{n+1})) = \sigma^2(x_{n+1}) = \sigma^2$ and $n$ is large enough.
At root, both point and interval predictors
are simply input-output relations with good stochastic properties. 

\subsection{Problem Classes}
\label{probclasses}

To provide an intuitive framework for thinking about the broad class of predictors
and predictive problems
we define three subclasses of problems roughly parallel
to those of \cite{Bernardo:Smith:2004} for model selection; one key difference is
that the second and third classes here are defined to be obviously disjoint.
Another variant is in \cite{Dawid_Vovk_Shafer}.
The problem classes are defined by the relative position of a given predictor versus a
`best' predictor when a best predictor exists. \footnotemark
The simplest is ${\cal{M}}$-closed.
In this case, there is a finite list of predictors ${\cal{P}} = \{ P_1, \ldots, P_K\}$ in which 
each $P_k$ may be a parametrized class of predictors.  One $P_k$ is true in the sense
that it gives asymptotically correct PI's.  This reflcts the fact that the asymptotically optimal
predictor might not be finite sample optimal.
In the IID case, this often follows from an `$MSE = {\sf bias}^2 + {\sf Var}/n$' argument when a simple preditor
can outperform a complex predictor for small $n$.
As discussed below, in many dependent cases, the true model is equivalent to the asymptotically
best predictor.  In this case, we can put a prior on the list of predictors that represents our
pre-experimental belief abut their correctness.  

\footnotetext{A variant on this classification of problems is 
${\cal{M}}$-mixed, see \cite{Pena:Walker:2001}, and combines
${\cal{M}}$-closed and ${\cal{M}}$-open in the sense of \cite{Bernardo:Smith:2004}.}

The next most complex is ${\cal{M}}$-complete:   This is the case where there is an optimal
predictor but it is inaccessible to us.  Again, we have a class ${\cal{P}}$ but none of them is assumed
asymptotically optimal. 
We may be able to find good approximations to the optimal predictor
at least under some restrictions,
and in principle, in some cases, be able to identify it as a limit.   However, we are not able to
specify it explicitly.  As a generality, this is the problem
class most people who gather data think of themselves as having and it fits, if uncomfortably,
with Box's dictum.   The inaccessibility of a predictor can mean one of two things:
Either it correspoinds to a model is just too complex for us to write down, perhaps even to conceptualize,
or it requires hopelessly massive amounts of data to approximate satisfactorily.     In this case,
a prior over predictors may correspond to a prior over models but they are outside the
collection of models we would regard as potentially `true'.  In this case, a prior
would not purely represent our pre-experimental information about the phenomenon.  However,
we could interpret it as a prior over ther actions the $P_k$'s represent as predictors.
Loosely, one
could regard ${\cal{M}}$-complete problems as signifying limits of a class
of predictors or models that are included.     The term model and predictor are not equivalent:
If a model exists, it must give testable predictions, i.e., convert to a predictor.  However,
there are predictors that do not convert to a model.  Hence, class of predictors is usually 
larger than the class of models for a given problem.

Some examples will
help.  Sometimes it's a question of scale and how much is included in the model.
We
can imagine genes and chromosomal proteins but the details of how they interact are
not really accessible to us even though we might believe they can be accurately modeled
if we leave out many interactions we want to approximate as having negligible importance.
A different problem occurs in design of experiments.    As we take more samples in any design,
it is possible that we have dependencies that increase rapidly so that everything we measure
is dependent in some way on everything else we measure and the only way to understand the
dependence would be to do another experiment introducing more dependencies we would have
to include.  In this case, there is a model, we can imagine it and perhaps write down key features
but data limitations will make it forever unobtainable.   In these cases, a prior would represent
our predisposition to predictors that we thought would provide good approximations to
an optimal predictor assuming it exists.

The most complex class is ${\cal{M}}$-open.  In this case, we do not assume there is a DG 
and the concept of a best predictor (or true model) does not make sense.     There need not
be a population much less a sampling rule. Thus, all we have is prediction,
the properties of our predictors, and properties of the data stream 
(such as number of distinct values, a running `median'
defined by the stream rather than by a distribution, etc.).  It is easiest to think of this data as
non-stochastic as much streaming data is.  Computer traffic data, online sales data, web-scraping data \footnotemark ,
and sensor data are standard examples.  Some of this data may in fact be stochastic but there is nothing
stable enough about it to permit inference so the concept of a true distribution is irrelevant.   
More formally,
if $Y_1 \sim P_1$ and $Y_2 \sim P_2$ and there is no relation between $P_1$ and $P_2$ 
it is not clear that $Y_1$ will help predict $Y_2$.  As a more complex example,
one can argue that, for instance, the Complete Works of
William Shakespeare regarded as a sequence of letters leads to an ${\cal{M}}$-open problem
in the sense that no
analysis, no matter how sophisticated, would enable someone to generate a 155th sonnet.
In this ase, as noted before, a predictor may be best regarded as an action in a decision theory problem and
the prior is a weight on the action indicating how likely it is to be optimal.   Even so, there are 
techniques (see Sec. \ref{streaming}) that can be used.

\footnotetext{The automated collection of massive amounts of observational data has become
common and it is an interesting debate whether this leads to ${\cal{M}}$-complete or -open problems.}

${\cal{M}}$-complete and ${\cal{M}}$-open problems differ mainly in whether an acceptable `belief predictor'
or `belief model' has been formulated.  Loosely a belief model is one that is not known to be wrong,
see Subsec. \ref{decisiontheory}. In an ${\cal{M}}$-complete problem, the analyst has formulated a belief model
but it is inaccessible or the analyst does not wish to use it for some reason.  
Thus, any formulated models or predictors are assessed in terms of their relationship to the belief model or predictor.
In an ${\cal{M}}$-open problem a belief model/predictor has not been formulated by
the analyst so any formulated models/predictors are compared to each other.   
Thus, it is possible that a belief model/predictor exists but has not been formulated
by one analyst and has been by another.  

As a generality, we don't expect techniques that will be good for one class of problem
to work well for another class of problems.  Obviously, if we have an $M$-complete problem
and we treat it as $M$-closed often we will end up with a biased predictor based on a
wrong model and
if we treat an $M$-open problem as $M$-complete, we may end up with an approximate
model that cannot be correct because there is no `true' model.   

There is a tension between predictors that are physical model based in the sense they
have components that are intended to be correlates with various aspects of physical reality and
more general predictors that may simply be procedures to provide a PI in response to
a collection of data and a new input.  Sometimes the former predictors are called explanatory-based and
the latter are called algorithmic. The tension arises from interpretability.  An example
will help:  The usual predictor from a Bayes model average (BMA) is optimal under squared error,
see \cite{Dempster:1975} Subsec. 2.3, and under relative entropy, see \cite{Raftery:Zheng:2003} Theorem 4,
which locally behaves like squared error.  This BMA
is a sum of predictors weighted by the posterior probabilities of their models that generated them.
Thus, if two models are incompatible, pehaps simply because they are different, it 
is unclear how to interpret the BMA physically.   Tension between predictors,  especially
optimal predictors that are rarely individual models, and their
interpretation is typical.  In practice, the usual case is that the interpretable predictor does
not perform as well as the optimal predictor, the exception being when the interpetable
predictor corresponds to a model that is extremely close to the true model, assuming it exists.
This latter case tends to be rare, especially for complex DG's.
Thus, the scientist must often choose a tradeoff between high interpretability and high predictive
success.
This tension is familiar from other settings:  A nonparametric model
will often have better predictive performance than a physical model at the cost
of interpetability.

It is commonly assumed that the true model, assuming it exists, generates the
best predictor i.e., we are in an ${\cal{M}}$-closed problem or in an ${\cal{M}}$-complete problem
where we're sure the model on which the predictor is based can be made arbitrarily close to the
true model.   These assumptions are often naive:
Consider a complex DG, perhaps one that corresponds to a deep learning network.   
If we make predictions using an estimate of that network it is easy for the variance term
in the mean squared error to be large and the bias term to be small.    However, it
is easy to imagine simpler models that have much reduced variance and only slightly
increased bias.  This is the same point as is often used to argue that Bayes estimators,
although biased, perform better than `best unbiased' estimators.   That is, 
predictors based on a simplification of a true but complex model
may perform better than predictors from the true model, at least pre-asymptotically.
A clear but different sort of example of when a  wrong model is better than a true model
is given in \cite{Clarke:Clarke:2018}, Chap. 10.

What can we say about when the best predictor corresponds to a true model?
First recall that
\begin{eqnarray}
E(Y \vert X) =  \arg \min_f E(Y - f(X))^2 .
\label{cdnlpred}
\end{eqnarray}
From \eqref{cdnlpred},  it is seen that when $Y$ really is a function of $X$ the true model
is best.  However, this neglects the difficulty of finding $f(X) = E(Y \vert X)$.  One of the
earliest results to address this is Theorem 2 in \cite{Rissanen:1984}.  This result
essentially shows that when
the true model is an $ARMA(p, q)$ process that the natural predictor from the true model
is optimal in a variance sense.  This is generalized in \cite{Skouras:Dawid:1998}
and \cite{Skouras:Dawid:1999} who relaxed the distributional assumptions and
the assumption of squared error.  The key point is that both plug-in and Bayes
predictors generally converge to the true distribution assuming it is available to them.

It is important to note that these results are asymptotic and the Bayes predictors
can be outperformed in small samples -- although the degre of outpeformance 
decreases to zero -- even when the DG's are not highly complex as
assumed in the case above.   This is seen for Shtarkov predictors that satisfy a
different and pointwise optimality criterion, see \cite{Clarke:2007}, as well as for a sort
of hybrid predictor that incorporates model uncertainty and invokes a sort of
minimax criterion, see \cite{Wong:Clarke:2004}.  In this context it should be noted that
there is usually a gap between a model-based predictor and an algorithmic based predictor
in favor of the algorithmic predictor because it typically has smaller bias being drawn from
a larger collection of predictors.    Thus,  a 
model based predictor can often be improved by
incorporating nonparametric or other algorithmic techniques, see \cite{Le:Clarke:2019},
and the reverse holds as well:  
An algorithmic predictor can often be related to a model-based predictor to
ask if the loss of predictive power is worth the gain in interpretability, see \cite{Le:Clarke:2022}.    Model averaging
versus model selection is a case in point:  Is the predictive gain from a model
average (often due to a reduction of bias) worth the loss of interpretability from 
not choosing a single model?

\section{Paradigms for Prediction}
\label{paradigms}

The goal of prediction is to make inferences on random quantities.  If the quantities are
not random, i.e., fixed, then the correct term is estimation.    In principle one could estimate 
a realized outcome $Y=y$, however this would only include the uncertainty of estimation;
it would neglect the uncertainty in the generation of $y$.
Forming a predictor for, say, $Y$, may require estimating a parameter.  However, the worth
of such an estimator is judged by its ability to improve the prediction of $Y$.  Conventional
optimality properties of estimators may or may not be relevant.

As a generality, prediction and estimation have different uncertainty quantifications and
whether a modeling or physical quantity is treated as stochastic or non-stochastic is an
aspect of modeling.   In all cases, it is important to have a clear, well-defined problem
so that predictor (and model, if appropriate)  interpretation and comparison will be precise 
enough to permit reasoned debate.  We emphasize that although they are related,
predictor/model uncertainty (whether for explanatory models or algorithmic ones)
is not the same as interpretability.

Next we provide an overview of four common paradigms for Bayesian prediction, roughly in 
historiacl order that also corresponds roughly in order of
most common usage.   There areat least four other Bayesian paradigms for prediction that 
have emerged more recently that we have not included
such as conformal prediction \cite{Shafer:Vovk:2008} and \cite{Bersson:Hoff:2022},  
information theoretic-prediction \cite{Kontkanen:etal:1997}, 
\cite{Merhav:Feder:1998} and \cite{Ebrahimi:etal:2010},
fiducial prediction, \cite{Dawid:Wang:1993}, and \cite{Wang:etal:2012},
and streaming data (${\cal{M}}$-open) prediction e.g., based on hash functions, see \cite{Bahri:etal:2018} and \cite{Satuluri:Parthasarathy:2012} for
techniques and \cite{Muthukrishnan:2005} for general background or other techniques that treat data as
a string rather then as stochastic.    Comprehensively covering 
Bayesian prediction is the work of many volumes not a single paper and some of these latter four
have not `caught on' widely -- at least as of this writing. 


\subsection{Forward Bayes Rule Prediction}
\label{Bayesrule}

This started out as a minimalist school of thought on prediction simply observing that
under Bayes rule the (posterior) predictive density is
\begin{eqnarray}
p(y_{n+1} \vert y^n) = \int_\Omega p(y_{n+1} \vert \theta) w(\theta \vert y^n) {\rm d} \theta .
\label{postpred1}
\end{eqnarray}
Implicit in this approach is that there is exactly one data set and all inferences are conditioned on
all data.  The opening example in Sec. \ref{Intro}  is an instance of this.
In \eqref{postpred1} $p(y \vert \theta)$ is a proposed model and one value of $\theta$ is assumed
true.  Pre-experimental beliefs represented by $w$ which may or may not be subjective.
Clearly, \eqref{postpred1} extends readily to having many parametric families each having a
within-model prior provided there is also an across model prior.  The corresponding expression
for $p(y_{n+1} \vert y^n)$ (in standard notation) becomes
\begin{eqnarray}
 \sum_{k=1}^K W(k \vert y^n) 
\int_{\Omega_k} p_k(y_{n+1} \vert \theta_k) w(\theta_k \vert y^n, k) {\rm d} \theta_k ,
\label{postpred3}
\end{eqnarray}
which is also a BMA under squared error.  Expression \eqref{postpred3}
extends to nonparametric Bayes contexts as well, see \cite{Clarke:Clarke:2018} Chap. 8.4
and more generally \cite{Muller:etal:2015} and \cite{Ghosal:VanderVaart:2017}.
One key benefit of this -- aside from simplicity -- is that Bayes rule prediction directly
provides an accurate quantification of
uncertainty.  Uncertainty is arguably more important for prediction than any other inference goal.

The direct use of Bayes rule is justified on multiple grounds.   The best known reasoning is
that Bayes rule is a fact from measure-theoretic probability.   Thus, the posterior is well-defined
and it makes good intuitive sense to derive inferences from it. 
A second argument for the use of Bayes rule comes from \cite{Zellner:1988}.  The idea is that
one can define a class of operations that process input information into output information
and that the optimal way to do this is via Bayes rule.  That is, 
Bayes rule is the optimal way to construct inferences.

A more formal approach to forward Bayes rule prediction follows from De Finetti's theorem,
see \cite{DeFinetti:1937}.  Given
an exchangeable probability measure $M=M_n$, there is a IID parametric family $P_\theta$ and
a prior $w(\cdot)$ so that for any $A = A_1 \times \cdots \times A_n \in  {\cal{B}} ({\cal{Y}}^n)$
\begin{eqnarray}
M_n(Y^n \in A ) = \int_\Omega \Pi_{i=1}^n P_\theta(Y_i \in A_i) w({\rm d}  \theta).
\label{DeFinetti}
\end{eqnarray}
Now, \eqref{DeFinetti} gives 
\begin{eqnarray}
P(Y_{n+1} \in  A_{n+1} \vert  y^n) = \int_\Omega P_\theta(A_{n+1}) w({\rm d} \theta \vert y^n).
\label{postpred4}
\end{eqnarray}
Choosing an IID family $P_\theta$ and a prior $w$ in \eqref{postpred4}
gives a version of \eqref{postpred1} and
as $n \rightarrow \infty$ the effect of $w$ decreases.

Implementation of this version of Bayesian prediction mainly requires the posterior density
or distribution.  These can often be found or approximated in closed form in simple problems,
e.g., in exponential families when equipped with any prior although conjugate priors
are the easiest.  Techniques for finding the posterior are well-known and include
MCMC, variational Bayes, and evolutionary computation.  In practice, this is the most commonly
encountered  version of Bayes prediction.

\subsection{Decision-Theoretic Bayes Prediction}
\label{decisiontheory}

This is probably the most popular conceptual approach to Bayes prediction, perhaps because the
most popular justifications of Bayesian thinking are decision-theoretic.   Indeed,  Savage's axioms
see \cite{Savage:1954} or more recently \cite{Abdellaoui:Wakker:2020} are often cited as
the `reason' to be a Bayesian.  A second construction of decision-theoretic Bayesian
arguments is developed in \cite{Bernardo:Smith:2004} and a third set of arguments
for decision-theoretic Bayes comes from the complete class theorems noted in
the introduction of this issue.  All three of these are usually phrased in estimation terms
but it is clear how they generalize to prediction.

Focusing on the reason to do specifically Bayesian {\it prediction}, one argument comes
from the reasoning in \cite{Aitchison:1975}; this was seen in the opening example
of Sec.  \ref{Intro}.   This is a decision-theoretic argument on the level of predictive densities.  
Conceptually this is similar to the asymptotic argument for BMA, see \cite{Raftery:Zheng:2003}, 
when the across model prior concentrates on one model.   Indeed, both of these rest on relative entropy
as a loss function.

To be more general, we recall the basic decision-theoretic framework for prediction
as outlined in \cite{Vehtari:Ojanen:2012} for example.  Let ${\cal{A}}$ be the action space,
i.e., the predictors $\hat{Y} = \hat{Y}({\cal{D}})$ 
we might use for a future value $Y \in {\cal{Y}}$.     The outcome $Y=y$ 
plays the role of the true state of nature.
Whether we believe we can identify it or not, let 
$L: {\cal{A}} \times {\cal{Y}} \longrightarrow \mathbb{R}$, with values $L(\hat{Y}, Y)$,
be a loss function that specifies the cost to us of making prediction $\hat{Y}$ when $Y=y$.
Finally, we must assume that we have a distribution (here denoted as a density for
convenience)
for $Y$, $p(y \vert {\cal{D}})$, 
i.e., a post-data probabilistic description of our beliefs about $Y$.   
Now we can write the optimal action as
\begin{eqnarray}
\hat{Y}_{\sf opt} = \arg  \min \int L(\hat{Y}, y) p(y \vert {\cal{D}}) {\rm d} y.
\label{argminpostrisk}
\end{eqnarray}
So, the argument for Bayes prediction is that the Bayes predictor
achieves the minimal posterior risk under $L$.  

Sometimes, $L$ is regarded as a `utility' function $u$.  Little harm is caused by thinking `$u = -L$'.
Typical choices for $L$ are relative entropy (sometimes called log-loss), squared
error, absolute error, and linex loss.
It is usually helpful to think about a 
class of loss functions even if this is difficult to use in practice.  

In \eqref{argminpostrisk}, both the prior and the model are implicit.  It would 
be more explicit to write $p(y \vert {\cal{D}})$ as the posterior from a prior
$w$ and a model ${\cal{M}} \equiv \{ P( A  \vert \theta) \vert A \in \sigma(Y) \}$,
e.g., as the posterior predictive $m_n$.
If this is done then ${\cal{M}}$ usually corresponds to a `belief' model.
A belief model ${\cal{M}}^*$ is a model that makes
$p(y \vert {\cal{D}}, {\cal{M}}^*)$ i.e., $p(y \vert {\cal{D}})$ based on ${\cal{M}}^*$,
a convincing representation of our beliefs 
about $Y$.  Hence, we may in principle use 
$p(y \vert {\cal{D}}, {\cal{M}}^*)$ in place of $p(y \vert {\cal{D}})$ in \eqref{argminpostrisk}.
In practice, belief models are rarely used because they are unavailable or unconvincing.
After all, if we have a belief model,  we can predict using \eqref{postpred2}.
See \cite{Vehtari:Ojanen:2012} for an alternative view which also provides an accessible and
comprehensive overview of model selection assessment in the decision-theoretic framework.
As a generality, it is not the specific examples of decision-theoretic predictors and assessments of
them that are useful
so much as the conceptual framework for thinking about prediction error as
an average loss.  This also permits including operations besides
integration in \eqref{argminpostrisk},  see \cite{Rostek:2010} or
\cite{Yu:Clarke:2011}.

\subsection{Reverse Bayes Rule Prediction}
\label{RBayesrule}

In Subsec. \ref{Bayesrule} we decomposed $M$ by De Finetti's theorem
to obtain conditionals for the $Y_{i}$'s
given the parameter and saw these could be used to form posterior predictives
$P( \cdot \vert y^n)$.
Here we do the reverse: We choose a sequence of posterior predictives directly.
Then we verify they combine to form a valid $M$. 
The key result is the Ionescu-Tulcea Extension Theorem.  This theorem states that
any sequence of conditional 
probabilities of the form $P_i(Y_{i+1} \in A_i \vert  y^i)$ for $i= 1, 2,\ldots , $ initialized 
by a given $P_1$ for $Y_1$ corresponds to a unique overall probability measure 
$P_\infty$ on the countable product space $({\cal{X}}^\infty, {\cal{S}}^\infty)$ that has the
prescribed conditionals.
For a formal statement see \cite{Shalizi:2007}.  Also, in the present context see
\cite{Fortini:etal:2000} for one of the earlier contributions as well as \cite{Berti:etal:2022}
and \cite{Fong:etal:2022} for more recent contributions.  If $Y_{n+1}, Y_{n+2}, \ldots$
are marginalized out of $P_\infty$ the result is $P_n$,  one choice for $M_n$ in
\eqref{DeFinetti}.

The De Finetti Theorem and the Ionescu-Tulcea Theorem are not converses but they
let us think interchangeably about sequences of
posterior predictive distributions, probabilities for the overall DG, and forward
Bayes rule prediction.  Moreover,
two benefits of modeling a DG by positing $P( \cdot \vert y^i)$'s for the $Y_{i+1}$'s
rather than positing $P_\theta$'s for the $Y_i$'s and a prior density $w$ to get $P$
are that $i)$ we bypass prior selection and $ii)$ we only assign probabilities
to observable quantities.


Bayesians, and other statisticians, are less familiar with modeling via sequences
of predictives than via conditional densities given a parameter.    However, it can be argued
that in practice reverse Bayes rule prediction and Bayes rule prediction are equally
effective as well as equally justified,  see \cite{Fong:etal:2022} and \cite{Berti:etal:2021}
amongst others.     Indeed, often, we do not recognize that we are doing reverse Bayes rule
prediction.  For instance,  a simple prediction rule like `tomorrow's daytime high $Y_{n+1}$
will be the same as today's daytime high $Y_n$' can be expressed as a reverse 
Bayes rule predictor by writing $Y_{n+1} \sim N(y_{n}, \hat{\sigma}^2)$ where 
$\hat{\sigma}$ is any realistic estimator of the variance; the normal is simply a convenient choice.
Ionescu-Tulcea guarantees there is a $P_\infty$ 
and De Finetti guarantees that $P_n$ can be decomposed into a mixture of IID conditionals,
even if we never bother to find $P_\infty$ or the its decomposition.   

The empirical distribution function (EDF) 
\begin{eqnarray}
\hat{F}_n(y) = \frac{1}{n} \sum_{i=1}^n I_{\{ y_i \leq y\}}
\label{EDF}
\end{eqnarray}
is another choice for $P_i(Y_{i+1} \in A_i \vert  y^i)$
and any density estimator can give a $p_i(Y_{i+1} \in A_i \vert  y^i)$.
Again Ionescu-Tulcea applies.
If the $Y_i$ are exchangeable more is true:   the De Finetti Theorem ensures that $\hat{F}_n$ and
$P_n(A) =  P( Y_{n+1} \in A \vert Y^n = y^n)$ converge weakly and a.s. to the same limiting 
marginal DF $F$, i.e.,  an $F$ that satisfies $Y_i \vert F \sim F$ where the $Y_i$'s are independently and
identically distributed (IID).   Formally, $F$ is the `model' (and is equipped with a prior).
Thus, the predictives in \eqref{EDF} converge to a limiting distribution.

The point is that, with some practice, one can develop a facility for proposing sequences
of predictives as readily as proposing parametric families. In reverse Bayes prediction 
it is these sequences that form the model.
In point of fact, reverse Bayes prediction is only `reversed' in contrast with the usual Bayes prediction.
A better term might be direct predictive modeling.

\subsection{Prequential Prediction}
\label{preqpred}

Prequential is a portmanteau of {\it pre}diction and se{\it quential}.
The founding document of prequential prediction is \cite{Dawid:1984},
but precursors are cited in the references.   For instance,
it was shown in \cite{Dawid:1982} that from within the Bayesian paradigm
bias is not detectable and one (partial) way around
this is via calibration because it ensures that the predictions of a probability 
forecasting system (i.e., a sequence of posterior predictives) will match the probabilities of
outcomes of the DG.  Unfortunately, as discussed earlier, calibration is often ineffective usually due
to insufficient data.   Another problem that can occur with calibration is that successive
recalibrations of a forecasting distribution, may violate the marginalization requirements
for the Kolmogorov Extension Theorem and hence not form a stochastic process.
One example of this is the Shtarkov predictor, see \cite{Shtarkov:1987}; see also
\cite{Rissanen:1996} and \cite{Clarke:2007}.)  

The prequential approach is an alternative way for the Bayesian to take bias into account.
The idea is to compare predictions (numerical, i.e., point, or probabilistic) directly to values of the predictand.
Pure prequential prediction assumes ordered data and invokes the `Prequential
Principle' that a predictor should be assessed only by comparing the sequence
of predictions it gave with the corresponding sequence of realized values and this
assessment should be independent of the way the predictions were constructed.  That is, 
assessment of a predictor sequence is based only on the
predictors $\hat{Y}_{i+1}$ or $\hat{P}(y_{i+1} \mid y^{i})$ and the 
realized values $y_i$ for $i=1, \ldots , n$. 
The point is that predictions that might have been made (but weren't) and values
of the predictand that might have been measured (but weren't) are irrelevant.
In common with the Likelihood Principle, prequential inferences depend only on the 
observed data and in common with reverse Bayes predition, prequentialists
only have to put distributions on observables.

The Prequential Principle ensures that predictive error depends only on the data,
the predictions, and that the evaluation of predictive errors is independent of the
way the predictions were generated.  This is a sort of fairness principle ensuring no predictor gets an
advantage from having been constructed using extra information such as aspects of the model
that may have generated the data. 
See \cite{Dawid:Vovk:1999} for a discussion of prequential
principles.

Prequential prediction,  in contrast to the earlier subsections, is a philosophy of prediction in general
rather than specifically Bayesian prediction.   Indeed, Bayes predictors (however derived)
are generally asymptotically prequentially optimal, but so are (frequentist) plug-in predictors in which the
only updating upon receipt of more data may be a parameter estimate.     That is,
Prequentialism is consistent with frequentism and with Bayesianism, and the arguments for being
Bayes are conceptually disjoint from the arguments for being Prequentialist. 

Prequential predictors can be either probabilistic or point predictors as discussed below
in Subsecs. \ref{probpreqpred} and \ref{pointpreqpred}.   Probabilistic prequential
predictors are of the form \eqref{cdldtn}:  They assign probabilities to future events based on 
past data and their errors 
are assessed sequentially usually by score functions that often compare data to probabilities.
Prequential point
prediction systems make predictions on the same scale as the predictand and are often 
assessed sequentially by loss functions.  Arguably, 
probabilistic prequential predictors 
are a neater fit with Bayesian than frequentist formulations. because Bayes rule concerns densities
or whole probability measures.  Also, in principle,    
prequential point predictors can be made comfortable with Bayes,
see \cite{Dawid:Vovk:1999} Secs. 6-7.

\subsubsection{Prequential probabilistic prediction}
\label{probpreqpred}

The prequential approach has two parts:  Construction of predictors and evaluation of predictors.
The concept of prediction is minimal:  There is no necessity for undergirding theory:
We look only at realized data and forecasts, not hypothetical repetitions,
what measurements we might have obtained but didn't, or even the provenance of the data.
The evaluation of predictors can be done in absolute terms (this predictor performs this well
in this sense) or relative terms (this predictor performs this much better than that predictor
in this sense).   Like forward Bayes prediction, the posterior predictive is an example
of a prequentially permitted forecasting system, see \eqref{postpred1} and \eqref{postpred2}.
Another common forecasting system is called plug-in.  Suppose $q_\theta(y) = q(y\mid \theta)$ is the probability 
density we will use to predict outcomes of a true probability $P$.  If we have an estimate
$\hat{\theta}$ for $\theta$ we can form
\begin{eqnarray}
q(x_{n+1} | \hat{\theta}(x^n))
\label{plugin}
\end{eqnarray}
and obtain predictors (interval or point) from it.  Indeed, any density valued function of
$x^n$, say $q(\cdot , x^n)$, is a prequentially
permitted predictor because part of the point of prequential prediction is to allow
reselection of a predictor via patterns of errors.  

Assessments of predictors are done by comparing forecast distributions to outcomes even 
though they are not on the same scale.    The idea is that the more representative an outcome
is to the forcast distribution, the better the distribution was.  Given $Q$ for continuous data $y^{n+1}$
with density $q$,
one common choice of cumulative assessment is the log-density $L(Q, y^{n+1})$
given by
\begin{eqnarray}
 -\log q(y^{n+1}) =- \sum_{i=0}^n  \log q(y_{i+1} | y^i),
\label{postpred3log}
\end{eqnarray}
and the closer the $y_i$'s are to the mode of $q$, the more representative of $q$ they are.
Expression \eqref{postpred3log} is just the (negative) log-likelihood that
 also has an interpretation in terms of codelegnth, one definition of information.  
Choosing $q$ to be the posterior predictive in \eqref{postpred1} gives $-\log m(y^{n+1})$.
In general, choosing \eqref{plugin} gives 
\begin{eqnarray}
L(Q,y^{n+1}) = -\log \Pi_{i=0}^{n} q(y_{n+1} | \hat{\theta}(y^n), y^n).
\label{pluginlog}
\end{eqnarray}
If we have two probability forecasting systems, say $Q$ and $R$, neither assumed true,
they can be compared analogously to in \eqref{postpred3log} by
\begin{eqnarray}
L(Q, R, y^{n+1}) = \log r(y^{n+1})/q(y^{n+1})
\label{comparelogpostpred}
\end{eqnarray}
and analogously to \eqref{pluginlog}, $L(Q, R, y^{n+1}) $ is
\begin{eqnarray}
\log \Pi_{i=0}^{n} r(y_{n+1} | \hat{\theta}(y^n), y^n)/\Pi_{i=0}^{n} q(y_{n+1} | \hat{\theta}(y^n), y^n).
\label{comparelogplugin}
\end{eqnarray}
None of these assessments makes any reference to a true DG, apart from the data.

Another way to assess a specific probabilistic predictor, say a distribution function also 
denoted $Q(Y_i \leq y | y^{i})$, is to transform it to the uniform.  That is, write
$U_i = Q(Y_i)$ so that $u^n$ is a vector of outcomes from the $Unif[0.1]$ if $Q$
is correct.  Now, tests of uniformity (and independence) can be done using $u^n$ and extends
to the non-dentical case, i.e., replace $Q$ for $Y_i$ with $Q_i$.

The log-loss, (or cost), is an example of a score function\footnotemark.   In general, a score function, denoted
$S(y, Q)$, is a way to assign a distance between data points $y$  and probabilities.   
There are many ways to construct score functions.  One is to start with a probability $Q$
and loss function $L$.  The risk for an action $a$ is $R(P, a) = E_P(Y, a)$.     Denote the
Bayes action by $a_Q = a_{B, Q}$.  Now, $S(y, Q) = L(y, a_Q)$ is a score function
and we can set $S(P, Q) = E_P S(Y, Q)$, implicitly assuming $P$ is true.
Now, $S(P,P) = E_P( S(X, P) ) = E_P L(Y, a_P) \leq E_P L(Y, a_Q)  = E_P S(Y, Q) = S(P, Q)$.
That is, the scoring rule $S$ is minimized by the true distribution and hence is proper.  If the inequality is strict
for $P \neq Q$ then $S$ is strictly proper; an example is the relative entropy.
Score functions can be used in place of log in \eqref{postpred3log}, \eqref{pluginlog},
\eqref{comparelogpostpred}, and \eqref{comparelogplugin} and for any other prequential predictor,
including those with explanatory variables.   Evaluation by
score functions is a natural add-on to reverse Bayes prediction since it is a method to
construct prequentially valid predictor sequences.

\footnotetext{Some authors define a score function, sometimes a scoring rule, as a function
comparing a realized outcome with its prediction.  We defer this to Subsec. \ref{pointpreqpred}
where we invoke a loss function or a negative utility function, suppressing the fine points of
terminology.}

Comprehensive overviews of these basic ideas that go beyond the
summary here can be found in
\cite{Dawid:2013} and \cite{Dawid:2018}.   These include concepts of prequential consistency, 
computability, and score function construction.
See \cite{Skouras:Dawid:1999} for prequential efficiency.   

Evaluation by comparing forecast disributions to outcomes is essential to probabilistic prequential prediction.
Score functions are only one way that has been studied, for more details
see \cite{Gneiting:Raftery:2007}, 
\cite{Gneiting:Katzfuss:2014} and 
\cite{Dawid:Musio:2015} amongst others.   Also,  an R package {\sf scoringRules} is also available,
see \cite{Jordan:etal:2018}.  However, the use of score functions has detractors such as
\cite{Brehmer:Strokorb:2019} who criticize their performance for tail values
or \cite{Clarke:Clarke:2018} Chap. 5.5 who argue they can be unstable, i.e., that different score functions 
can vary greatly in their inferences while individual score functions can be insensitive to the
class of $Q$.   The intuition is that since score functions actually operate on the level of densities
and densities are more distant from the data than distributions are, the link between the
scores and predictions will rarely be robust.  Issues relating score functions to sensistivity are discussed in detail in \cite{Fissler:Presenti:2023}.

\subsubsection{Prequential point prediction}
\label{pointpreqpred}

While many prefer probabilistic forecasts, point predictors are at least as important.  For instance,
it is one thing to know the predictive distribution of lifespan; it is another to say a your median
lifespan is predicted to be so many years.  Also, in many real world cases, a probability forecast 
cannot be implemented, a number must be chosen.   There is real benefit for the predictor to
be in the same class as the outcome and point predictors can be efficient, \cite{Skouras:Dawid:1998}.

It is well known that when explanatory variables are present, multiple models
may be indistinguishable numerically on a restricted doman up to a limit determined by the sample size.
In this case, it is obviously enough to ignore concerns about model uncertainty and focus on the
predictions only.  While this can be done in the context of Subsec. \ref{probpreqpred}, is is easier
with point predictors.

Most treatments of prequential point prediction adopt a de facto decision theoretic formulation,
regarding the score function as a loss function.
A minimal criterion for a good predictor to satisfy would be 
that the predictions be close to the outcomes.
So, suppose we have a point predictor $\hat{Y}$ as before and a loss function $L$ 
so that the predictive error (PE) between
$\hat{Y}_{n+1}(x_{n+1})$ and $y_{n+1}$ is
\begin{eqnarray}
PE = L(\hat{Y}_{n+1}(x_{n+1}), y_{n+1} ) 
\label{PE}
\end{eqnarray}
where 
$y_{n+1} = y_{n+1}(x_{n+1})$.    If we make $n$ predictions in sequence then
we also have a cumulative predictive error (CPE) that may serve as a better
assessment of a predictor along a sequence.  That is,
\begin{eqnarray}
CPE = \frac{1}{n} \sum_{i=1}^n L(\hat{Y}_{i+1}(x_{i+1}), y_{i+1} ). 
\label{CPE} 
\end{eqnarray}
We want the CPE to be small, as a way to make the PE small.
The question of how small depends on the class
of prediction problems we have.  For $M$-closed 
we can ask that, say, the CPE be minimal, asymptotically in $n$.   
For ${\cal{M}}$-complete problems we can ask that the CPE be minimal,
asymptotically in $n$ as as approximation to the true model converges
to the true model.  For ${\cal{M}}$-open problems, there is no minimum in general.
So,  we invoke other properties of predictors,
see Sec. \ref{streaming}.   Some formally ${\cal{M}}$-complete problems may be so difficult tht
we may be forced 
to act as if they were ${\cal{M}}$-open.  One benefit of using point predictors is that
\eqref{PE} and \eqref{CPE} easily make sense for all three classes of problems; in ${\cal{M}}$-open
problems it is difficult to relate a probability forecast to an outcome for which there is no model. 

For $M$-closed and $M$-complete problems
where it is possible to generate a meaningful PI, we can ask how often the outcome lands
in the PI.  Thus, we may choose the $L$ in \eqref{CPE} to be
$$
L( PI(x_{n+1}), y_{n+1} ) = \begin{cases} 1 & y_{n+1} \in PI_{n+1} \\
0 &  y_{n+1} \notin PI_{n+1}
\end{cases},
$$
essentially zero-one loss on whether the outcome was in the PI at stage $n$.
Other choices of $L$ are possible e.g., if $\mu_{n+1}$ is the midpoint of $PI(x_{n+1})$,
we could choose $(\mu_{n+1} - y_{n+1})^2$ for $y_{n+1} \in PI$ and some larger fixed
positive number otherwise.

A problem that has not been discussed heretofore is updating the predictor or
the predictive structure more generally.  This continues the comment at the end
of Subsec. \ref{RBayesrule}.
In many settings even if we have $n$ data points we expect to refine our
predictor after receiving more data.  
So far, this has only been seen by re-updating a posterior predictive (Subsec. \ref{Bayesrule})
and the plugin case (Subsec.   \ref{probpreqpred}).  Both of these are minimal updates
of a predictor thought to be basically acceptable.  

More generally the prequential strucure permits reformulating the predictive peoblem 
entirely in response to data.  
As an example, suppose in an ${\cal{M}}$-complete or
${\cal{M}}$-open problem we have used a relevance vector machine (RVM), see \cite{Tipping:2001}.
The RVM formed from the first $n$ data points may be deemed acceptable for, say,
the next $k$ data points but the $n+k+1$ value of $PE$ or $CPE$ may be too large forcing us to
rederive the RVM using the $n+k+1$ data points.  The result would very likely be a
different RVM.  So, in these complex problems
we want a rule that tells us when to choose a new predictor, not necessarily from the same class.  

In fact, prequentialism permits us to rethink the entire prediction problem including
the loss and the model class from whch predictors are drawn in response to the PE, CPE, or other criteria
that satisfy the prequential principle, e.g., robustness.
Following  \cite{Dawid:1992} who used multiple examples, we may be led to sequential 
reselection of the prediction problem, e.g., the autoregressive order in a time series,
even before estimating any parameters.
This is analogous to rechoosing a regression model
based on the residual errors.  Such approaches have been used to good effect 
most notably in \cite{Erven:etal:2012}; but see also \cite{Clarke:Clarke:2009}
and \cite{Wong:Clarke:2004}.  These results suggest prequentialism may achieve a
better trade off between model list bias (how well the models on a model list reflect the DG)
and model list variability (how much variability is contributed by the differences across models
on a list) where the data is complex.

\subsection{Notes to Sec. \ref{paradigms}}
\label{notes1}

The predictive paradigm in statistics has a long history.   Some classics are
\cite{Aitchison:Dunsmore:1975} and \cite{Geisser:1993}.  More recently are
the books \cite{Meeker:etal:2017}, \cite{Clarke:Clarke:2018}, and the popularizations
\cite{Silver:2012} and \cite{Siegel:2016}.  There are many others, but they do not
take prediction as the central organizing principle of Statistics.

As presented here, the field of Bayesian prediction has been defined as a $2 \times 3 \times 4$ table
for problems:  model based vs algorithmic;  ${\cal{M}}$-closed, -complete, -open; and 
Bayes rule, decision theoretic, reverse Bayes rule, and prequential.
If the extra four Bayes predictive philosophies are included we get a $2 \times 3 \times 8$
table.

Very loosely, these four philosophies are as follows.  {\it Conformal:}
The key principle is that future data will look a lot like past data.  To this end a
noncoformity measure is introduced and used to define a prediction region 
that has high probability, essentially in the \footnotemark EDF.    {\it Information theoretic:}
Probabilities are regarded as codelength functions, a form of data complexity.  Then,
complexity is minimized in some formal sense to obtain a predictive distribution.  
The relative entropy argument given at the beginning of Sec.  \ref{Intro} can be seen 
as an example of this, but direct usage of the (explicit) Shannon code length function
essentially an entropy or the Shannon mutual information -- regarded as a measure
of dependence -- between past and future data is more typical.
{\it Fiducial:}   Suppose $Y= G(\xi, U)$ where $G$ is known where $U$ is a random variable
and $\xi$ is a parameter.  Given $Y=y$, invert it by writing $Q(y,u) = \{\xi | G(\xi, u) = y\}$.
This gives $Q(y, U^*)$ where $U$ is an independent copy of $U$ that can be used to
give prediction.  Finally, {\it Streaming data:} For data to be analyzed in one pass, one idea
is to generate random hash functions i.e., functions that are many to one, that
give estimates of streaming analogs of ${\cal{M}}$-closed quantities such as means and medians.
In this case, the randomness is in the hash functions, not the data.  We defer this and related
technqiues to Subsec. \ref{hash}.

\footnotetext{The empirical distribution function (EDF) is good because it satisfies a large 
deviation principle, the Kiefer-Wolfowitz theorem.}

\section{Unitary Predictor Selection}
\label{Predsel}

Henceforth we distinguish between unitary and composite predictors.  The idea is that a unitary predictor 
corresponds to one model or other single entity that generates a prediction.  A composite predictor can be
decomposed into multiple unitary predictors that do not lend themselves to further division.
Usually, unitary predictors correspond to a model, often an explanatory model,  but this is not
necessary because the set of all predictors that come from models (explanatory or algorithmic)
is a subset of all predictors.
In the specific case of a unitary explanatory predictor that comes from a model, there are components that correlate
to physical quantities.   Choosing a model may involve specifying these correlates
and their interactions on the basis of assumed physical knowledge.   If so, then the individual components
may be subject to validation.    We do not consider that case here; it is to be answered by experimentation
or robustness criteria beyond our present scope.
Thus, we implcitly assume that our unitary predictors (or models) are algorithmic or regard explanatory models
as if they were algorithmic ignoring their physical basis.  Moreover,
composite predictors -- that may sometimes be regarded as composite models (explanatory or algorithmnic)
even though this is
not the typical case -- are harder to evaluate since physical correlates are harder to examine individually.
So as a rule, we treat the composite predictor as an entity unto itself i.e., we rarely assess the individual
predictors or  models as unitary apart from how they affect the overall predictions.

In this section, we look at selection for unitary algorithmic predictors.  There are two cases:
1) The selection criterion itself is (algorithmic) model based subject to the
condition that the model converts to a unique, tractable predictor that makes testable predictions, or,
2) The selection criterion itself is predictive and likewise yields a tractable predictor that makes testable predictions.
The distinction is, roughly, between the concepts of model fit on existing data
and model validation on `future' data not used in the analysis.
The first is better developed and more familiar
than the second which is more recent and often harder.
In using individual models for prediction we have two tasks, choosing the model and hence the predictor
and then evaluating how well it has done.  Thus, predictive assessment 
may play two roles; one in predictor selection and one in predictor validation.

The intuition is that a predictor selected on the basis of a predictive
criterion should be good predictively and one that is validated predictively post-data
should be good pre-data predictively.    Obviously, non-predictive criteria for predictor selection 
may also perform well.    Here, it is enough to focus on predictor selection because we
have already discussed predictor validation -- the conceptual advance over earlier inferential methodologies
that the predictive approach represents 
-- at multiple points earlier, e.g., using loss functions, CPE,
score functions, etc.  The disjunction between predictor selection and validation is the key to
prequentialism.  In Sec. \ref{ModelAverages},  we look at composite predictors.
We note that selection for non-model based (algorithmic or explanatory) predictors seems to be a
gap, perhaps partaiully because the class of such predictors is not well-defined.

Recall that we consider models $p_k(y \vert x, \theta_k)$ having within-model priors $w_k(\theta_k)$
and an across-models prior $W(k)$ for $k= 1, \ldots, K$ where we allow $K = \#(\mathbb{N})$,
at least conceptually.  The subscript $k$ on $p_k$ allows for
different subvectors of $x$ to be chosen.  We use $M = M_k$ generically to mean $\{ p_k, w_k\}$.

\subsection{Predictive Predictor Selection}
\label{nonpredsel}

There are many ways to select predictors or models from either a Bayes or non-Bayes standpoint
and the focus here is on prediction so we discuss only a few.  We recall that
many non-Bayes methods work well -- although probably for Bayes reasons e.g.,
the complete class theorem, as discussed in the opening editorial.    So, while our
focus is on Bayesian predictive predictor selection, we include discussion of some
popular non-Bayes methods for predictive predictor selection.

\subsubsection{Posterior predictive calibration (PPC)}

Under the containment principle, parameters, observations, and future observations have a joint
distribution ona  single measure space.   Also, a
prediction is calibrated if probability statements about parameters, observations,  and
future unseen data are in agreement with repeated experience.  Calibration is usually assessed in terms
of the frequentist concept of probability as a long run average of occurences of events, e.g.,  a model is
calibrated if it outputs 95\% confidence intervals for an unknown quantity of interest $\mu$
so that 95\% of them contain $\mu$ under repeated realizations.
A correct Bayesian model is always calibrated because $W(\mu \in I \vert {\cal{D}}) = E_m W( \mu \in I \vert {\cal{D}}) = .95$.
The analogous predictive statement is
\begin{eqnarray}
\nonumber
&& M(Y_{n+1} \in I_{.95}(y^n) \vert  y^n) \\
&& = \int_{I_{.95}(y^n)} w(\theta \vert y^n) p(y_{n+1} \vert \theta) {\rm d} \theta {\rm d} y_{n+1} =.95.
\label{ppc1}
\nonumber
\end{eqnarray}
That is, over repeated future sampling of parameters and future outcomes, the 95\% 
PI averages out to 95\% in the posterior predictive distribution.

Given this, we can use a posterior predictive $p$-value to check our calibration and
hence whether a belief model is adequate or should be replaced.  Alternatively,  we can restrict attention
to belief models that are, with respect to future data, properly calibrated given our observations.

The idea is to choose a test statistic $T(y, \theta)$ that is a function of both the data and parameter.
We first obtain $w(\theta \vert y^n)$ typically via MCMC draws $\theta_1, \ldots, \theta_S$.  Given
$\theta_s$ we draw $y^{{\sf rep}, s}$ from $p(y \vert \theta_s)$.  Now we have
$T(y^{{\sf rep}, s}, \theta_s)$ and $T(y^n, \theta_s)$.

If the realized test statistics $T(y^n, \theta_s)$ seem atypical in the posterior predictive distribution
of $T(Y^{\sf rep}, \theta)$ then we conculde the model and hence its predictor are discredited.
That is, if the posterior predictive $p$-value 
$$
\hat{M}( T(y^{\sf rep}, \theta) > T(y^n, \theta) \vert y^n ) 
\approx \frac{1}{S} \sum_{s=1}^S 1_{T(y^{{\sf rep}, s}, \theta) > T(y^n, \theta)}
$$ 
is too close to 1 or 0, we conclude that the model is wrong.
Perhaps the largest drawback of PPC is that the PPC p-value is not uniformly distributed under
the null hypothesis that the model is is true (often peakier), making hypothesis testing cumbersome.

\subsubsection{Optimizing a functional.}

Let $U$ define a functional that evalautes the `worth' of a model.
That is, $U$ may be a utility function, loss function, score function, or other sense
of distance, that assesses how good a model is.    Given ${\cal{D}}_n$ let ${\cal{D}}_n^*$ be an
idential second copy
assumed to come from the same DG.   One post-data predictive approach is to
assign worth to a candidate model $M$ by
\begin{eqnarray}
W(M \vert x) = \int [ U \circ p(y \vert x, {\cal{D}}, M)  ] p_T( y \vert x)  {\rm d} y
\label{Worth1}
\end{eqnarray}
assuming $p_T$ exists or is replaced by a `belief model'.  The empirical value
of \eqref{Worth1} is
\begin{eqnarray}
W_{\sf emp} (M \vert {\cal{D}}) = \frac{1}{n} \sum_{i=1}^n [U \circ p(y_i^* \vert x_i^*, {\cal{D}}, M)].
\label{Worth2}
\end{eqnarray}
Sometimes this is called the posterior predictive approach even though the posterior does not explicitly appear.   The term
predictive is valid because a possible future value of $Y$, denoted $y$ or $y_i^*$, at $x$ or $x_i^*$, is used.
So, these quantities are not assessing fit, at least not directly.
If $U = \log$ then \eqref{Worth1} and \eqref{Worth2} are clear  If, say, $U$ is squared error then
$$
(y_i^* - E(Y \vert x_i^*, {\cal{D}}, M))^2,
$$
where
$$
p(y \vert x_i^*, {\cal{D}}, M) = \int p(y \vert x_i^*, \theta, M)  w(\theta \vert {\cal{D}}, M) {\rm d} \theta,
$$
is used in the bracketed expressions in \eqref{Worth1} and \eqref{Worth2}.

More complicated is a conditional predictive entropy form of \eqref{Worth1} given by setting $W(M \vert x)$ to be
\begin{eqnarray}
 \int \left( \int p(y \vert x, \theta, M) \log  p(y \vert x, \theta, M)   {\rm d} y \right) w(\theta \vert {\cal{D}}, M)  {\rm d} \theta.
\label{Worth3}
\nonumber
\end{eqnarray}
There are many other examples.

Now, choosing $M$ to achieve
$$
\arg \min_k W_{\sf emp} (M_k \vert {\cal{D}})
$$
may give a model that gives good predictions.  However, the main drawback is
that using the data twice may amount to over-fitting because ${\cal{D}}^*$ is only
nominally not the same as ${\cal{D}}$. That is, \eqref{Worth2}
is a biased estimator for \eqref{Worth1}, although likely the bias $\rightarrow 0$ as $n \rightarrow \infty$.

\subsubsection{Using a Functional and a hold-out set.}

To reduce the effect of double use of the data, some predictive criteria split the data as
${\cal{D}} = {\cal{D}}_1 \dot{\cup} {\cal{D}}_2$.  Often ${\cal{D}}_1 = {\cal{D}}_{\righthalfcap i}$ for some $i$
where $\righthalfcap i$ (`not $i$') indicates the complement of $(x_i,y_i)$ in ${\cal{D}}$.
With this, \eqref{Worth2} can be modified to
\begin{eqnarray}
W_{\sf emp} (M \vert {\cal{D}}) = \frac{1}{n} \sum_{i=1}^n [U \circ p(y_i \vert x_i, {\cal{D}}_{\righthalfcap i}, M)]
\label{Worth4}
\end{eqnarray}
with correspondingly modified forms for $\log$, squared error and other $U$'s.
This criterion is predictive in that ${\cal{D}}_{\righthalfcap i}$ enables the prediction of $y_i$ at $x_i$
but this also reflects fit since we are using the data we actually have not even treating it as nominally distinct.

Expression \eqref{Worth4} can be modified to allow for $K$-fold cross-validation (CV).
Simply replace the ${\cal{D}}_{\righthalfcap i}$'s with $\righthalfcap {\cal{D}}_k$'s for $k=1, \ldots, K$
where the ${\cal{D}}_k$'s are disjoint, equally sized ($n/K$) sets drawn at random from ${\cal{D}}$.
The sum in \eqref{Worth3} is then taken ensuring that in term $i$,  $(x_i, y_i) \in  {\cal{D}}_k$.
Again, this is a mix of fit and prediction akin to regular formulations of CV and the predictive error
sum of squares, the PRESS statistic, see \cite{Alcantara:etal:2023} for a recent variation.
Note that post-date predictive model selection usually over-states the worth of a model due to the
re-use of the data.  Using a hold-out set improves this problem but doesn't eliminate it because
there is still some re-use of the data.

When $U$ is $\log$,  \eqref{Worth1} and \eqref{Worth2} become
\begin{eqnarray}
\nonumber
W(M \vert x) &=& \int [ \log  p(y \vert x, {\cal{D}}, M)  ] p_T( y \vert x)  {\rm d} y \\ 
\nonumber
W_{\sf emp} (M \vert {\cal{D}}) &=& \frac{1}{n} \sum_{i=1}^n [ \log p(y_i^* \vert x_i^*, {\cal{D}}, M)]
\label{Worth5}
\end{eqnarray}
and are called the expected log posterior density of a model $M$ (ELPD), 
with a corresponding version for \eqref{Worth5}.
In practice, the ELPD is often calculated by
\begin{eqnarray}
\widehat{ELPD} =\frac{1}{n} \sum_{i=1}^n \log  \sum_{j=1}^J p(y_i \vert \theta_j)
\label{Worth6}
\end{eqnarray}
where $\theta_1, \ldots , \theta_J$ is a selection of points in the parameter space.
Analogs of \eqref{Worth6} such as
\begin{eqnarray}
\widehat{ELPD}_{CV} = \sum_{i=1}^n  \log p(y_i \vert {\cal{D}}_{\righthalfcap i})
\nonumber
\end{eqnarray}
under different $U$'s are also feasible.

These techniques are based on conditional distributions for a future outcome given
past outcomes and are for choosing a model.  In many cases, the desire is not just to choose a
model but to choose a specific density within the model i.e., estimate $\theta$.  In these cases,
it is common to use an MLE or other optimal value for $\theta$.  An alternative is to
invoke a `belief model' $\tilde{M}$ that does not make heavy assumptions such as a Bayes model
average (see next section) and minimze  quantity such as
\begin{eqnarray}
\int \log \left( \frac{p(y^n \vert x^{n,*}, {\cal{D}},  \tilde{M} )}{p(y^n \vert x^{n,*}, {\cal{D}},  \theta_k)} \right)
{p(y^n \vert x^{n,*}, {\cal{D}},  \tilde{M} )} {\rm d} y^n
\nonumber
\label{Worth7}
\end{eqnarray}
over $k$ and $\theta_k$, with corresponding expressions when a hold-out
set is used.

\subsubsection{Projection methods}

This class of techniques for finding a preditor relies on having a belief model.    There are many
ways to formulate a belief model; often these are taken to be model averages, see Sec. \ref{ModelAverages},
and called reference models.    If $\tilde{M}$ is a reference model with density 
$p( y_{n+1} \vert x_{n+1}, \tilde{\theta}, \tilde{M})$, it can be `projected' to the model
$p( y_{n+1} \vert x_{n+1}, \theta_k, M_k)$ by finding $\theta_{{\sf opt},k} $ defined by
\begin{eqnarray}
\label{projection1}
\arg \min_{\theta_k} \frac{1}{n} \sum_{i=1}^n U(p( y_{n+1} \vert x_{n+1}, \tilde{\theta}, \tilde{M}),  p( y_{n+1} \vert x_{n+1}, \theta_k, M_k)), \hspace{-60pt}
\end{eqnarray}
where $U$ is often taken to be the relative entropy which is known
to permit useful concepts of projection in convex spaces of densities.
The distance between $\tilde{M}$ and $M_k$ is
\begin{eqnarray}
\frac{1}{n} \sum_{i=1}^n E U(p( y_{n+1} \vert x_{n+1}, \tilde{\theta}, \tilde{M}),  
p( y_{n+1} \vert x_{n+1}, \theta_{{\sf opt},k}, M_k)),
\label{projection2}
\end{eqnarray}
where a minimization over $k$ is also possible.  These expression only make sense for the ${\cal{M}}$-closed
and -complete settings, but this may enable better variable selection.

Expressions \eqref{projection1} are only avalaible analytically in special cases
such as linear models and generalized linear models.  More generally, \cite{Dupuis:Robert:2003}
proposed generating values $\tilde{\theta}_s$, for $s=1, \ldots, S$
from the reference posterior and approximating \eqref{projection2} by
\begin{eqnarray}
\frac{1}{nS} \sum_{i=1}^n \sum_{s=1}^S 
U(p( y_{n+1} \vert x_{n+1}, \tilde{\theta}_s, \tilde{M}),  p( y_{n+1} \vert x_{n+1}, \theta_{k,s},  M_k)).
\label{projection3}
\end{eqnarray}
Then, the `projected' samples $\tilde{\theta}_s$ can be used for posterior inference by finding
the predictive density
\begin{eqnarray}
p(y_{n+1} \vert x_{n+1}, {\cal{D}}, M) = \frac{1}{S} \sum_{s=1}^S  p(y_{n+1} \vert x_{n+1},  \tilde{\theta}_s, M)
\end{eqnarray}
where $M$ is now the best choice from the $M_k$'s.  These techniques are described and developed in
\cite{Goutis:Robert:1998} and \cite{Piironen:Vehtari:2016} 
but do not seem to be used extensively yet.

\subsection{Non-Predictive Predictor Selection}

Again, predictor selection segregates into Bayes and non-Bayes.  Here we look
at non-predictive methods.  These are model based and,
 the data is mainly used for model fit rather than outcome prediction (or retrodiction).
The hope -- and it is only a hope -- is that a model that fits well will predict well.

\subsubsection{Testing for a Predictor:  Bayes Factors}

The Bayes factor (BF) is the ratio of two marginal likelihoods.  For model $M_1$ and $M_2$ it is
\begin{eqnarray}
BF_{12} = \frac{p(x \vert M_1)}{p(x \vert M_2)} &=& \frac{ \int w(\theta_1 \vert M_1)p({\cal{D}} \vert \theta_1, M_1) {\rm d} \theta_1}
{ \int w(\theta_2 \vert M_2)p({\cal{D}} \vert \theta_12 M_2) {\rm d} \theta_2}
\nonumber \\
&=& 
\frac{W(M_1 \vert {\cal{D}} )W(M_1)}{W(M_2 \vert {\cal{D}} )W(M_2)},
\label{BF1}
\end{eqnarray}
with high values favoring $M_1$.
Using BF's for hypothesis testing is optimal under generalized zero-one loss where the
generalization means that the `1' is replaced by constants $c_1$ and $c_2$
for the cost of the two sorts of errors.   Formally,  $c_1$ and $c_2$ are specified in advance of data collection
because they determine the threshold for
decision making.  In practice,  however,  the threshold is usually chosen post-hoc and separately from the loss.
Being ratios, BF's can be unstable especially when $n$ is small or
the prior is diffuse.  When $M_1$ and $M_2$ (or their priors) are too similar, BF's may be indeterminate.
For a lucid overview see \cite{Berger:Pericchi:2001}.

In the absence of explanatory variables, the model, or its predictor,
maximizing the BF can be used.    In regular hypothesis testing, the size of the BF determines whether
a hypothesis is accepted or rejected whereas in the predictive setting, the threshold doesn't
matter:  A predictor must be chosen and the one achieving the highest BF relative to other
candidate models is simply the logical choice.  The prior has a major effect on the BF
but prior selection is beyond our present scope.   Nevertheless, typical choices are
uniform, Jeffreys, and reference priors.  Sometimes these are called within-model priors.
When improper prios are used the marginal likelihoods may be indeterminate.  As a generality,
prior sensitivity analyses of the marginal likelihood and BF, 
especially for across model priors, are especially important in model selection.
Marginal likelihoods depend on the prior.  In model
selection for predictors diffuse priors can be actually very informative.
This is different from within model priors for parameter estimation and worse when explantory
variables are included. 

In the presence of explanatory variables, the selection of the prior becomes a bigger issue,
but the mechanics of BF's are the same.    Recall  \eqref{postpred3} for the typical form
of a marginal likelihood in this setting.    One way to see this in more detail is in the
context of linear regression.  Let $Y = X \beta + \epsilon$ with the usual assumptions
and write $\gamma = (\gamma_1, \ldots , \gamma_d) \in \{0,1\}^d$.    The class of all linear submodels is now
\begin{eqnarray}
Y = X_\gamma \beta_\gamma + \epsilon
\label{lingamma1}
\nonumber
\end{eqnarray}
as $\gamma$ ranges over $\{0,1\}^d$ and $X_\gamma$, $\beta_\gamma$ indicate the selection
of variables and parameters.  When $\gamma_j = 0, 1$ for exclusion or inclusion of $x_j$
respectively.  Cleary $\gamma_{\sf null} = (0, \dots , 0)$ means the modelwith
no explanatory variables and $\gamma = (1, \ldots , 1)$ measn the model with all
$d$ explanatory variables included.   Now, given across model and within-model
priors any BF of the form
\begin {eqnarray}
BF = \frac{p(y \vert M_\gamma)}{p(y \vert M_\gamma^\prime)}
\label{lingamma2}
\end{eqnarray}
when $\gamma \neq \gamma^\prime$ can be evaluated as a way to choose a model to generate a predictor.
Choosing $\gamma^\prime = \underbar{0}, \underbar{1}$ often give useful comparisons
and neighborhoods of models based on having a selection of explanatory variables that
are not too different are also possible.  Also, see \cite{Liang:etal:2008} for forms of
\eqref{lingamma2} for $g$-priors.

Many choices for within-models and across-models priors have 
been proposed.   Across model priors include the uniform prior, priors that are constants
over sets of explanatory variables of the same size, and Zellner $g$-priors (often with Jefferies
prior $\propto 1/\sigma^2$ on $\sigma^2$ and a flat prior on $\beta_0$).  Each has its
benefits and shortcomings but these are beyond our present scope but we note that
priors that penalize complexity are often favored.

\subsubsection{Variations on BF's}
\label{BFvariations}

It is well-known that the posterior will typically asymptotically assign probability one to the
model closest in relative entropy to the true model, if it exists.   However, few models are
uniformly good over the range of the $d$-dimensional covariate space.  So, to alleviate
convergence to an inadquate but best model, Bayes testing can be used to provide an
improved local approximation to the true model.  
One way to do this in the ${\cal{M}}$-complete
setting is \cite{Yu:etal:2013}.    Let ${\cal{D}}^* \subset {\cal{D}}$ be a set of data reserved 
to update models.
Partition the covariate space as $\mathbb{R}^d = \dot{\cup}_{k=1}^K \Lambda_k$
and write ${\cal{D}}_k = ( {\cal{D}}^*)^c \cup \{(y_i, x_i) \vert x_i \in \Lambda_k\}$.
The local BF is
\begin{eqnarray}
LBF_{1 2\vert k }= \frac{ p({\cal{D}}_k \vert M_1, {\cal{D}}^*)}{ p({\cal{D}}_k \vert M_2, {\cal{D}}^*)}
\label{PBF1}
\nonumber
\end{eqnarray}
where there is a tradeoff between the data used to update the posterior and the data
used to define the $\Lambda_k$'s.    Note that the  definition of the LDF holds for
parametric and nonparmetric priors.

If a model $Y = g(x) + \epsilon$ holds, then the procedure is start by fitting $J$ parametric models and 
assigning priors to each.   Then the data is paritioned into $J$ subsets and posteriors built by
conditioning on each subset indvidually.  These give marginal models that can be used to
form BF's using marginals on the compelements of the subsets for each $j$.
Each subset can then be clustered by CART and
the $\Lambda_k$'s corresponding to the clusters.  That is, 
the clustering gives a partition that leads to the $\Lambda_k$'s.
LBF's are then used to find weights for the $J$ procedures on each of the $K$ regions
and this leads to a weighted average of localized predictions from the $J$ models.
In many respects this is a variation on Bayes model averaging,  cf. \eqref{postpred3} and see  Sec. \ref{BMA}.

Three other forms of BF's occur for model selection, although none are well-studied.  The first 
are called intrinsic BF's, see \cite{Berger:Pericchi:1996}.   They are ratios of marginal likelihoods of the form
\begin{eqnarray}
IBF_{12} = 
\frac{p({\cal{D}}_{\sf min}^c  \vert M_1, {\cal{D}}_{\sf min} ) }{p({\cal{D}}_{\sf min}^c \vert M_2, {\cal{D}}_{\sf min} ) }
\label{IBF}
\end{eqnarray}
where ${\cal{D}}_{\sf min}$ is a minimal training set, i.e.,  a smallest subset of ${\cal{D}}$
that ensures all the posteriors formed from $w_k$'s are proper.  The benefit of this approach is
that it allows a wider range of priors to be used, namely those improper priors that
give proper posteriors for all $M_k$'s using a finite amount of data, independent of the full sample.
Other versions of ISI such as geometric, arithmetic, and median are also well-defined but not often used.

Another form of the BF is the fractional BF:
\begin{eqnarray}
FBF_{12} = \frac{q({\cal{D}} \vert M_1, f)}{q({\cal{D}} \vert M_2, f)}
\nonumber
\end{eqnarray}
where
$$
q({\cal{D}} \vert M_j, f) = \frac{ \int w_j(\theta_j ) p({\cal{D}} \vert \theta_j, M_j) {\rm d} \theta_j}
{ \int w_j(\theta_j ) p({\cal{D}} \vert \theta_j, M_j)^f {\rm d} \theta_j}
$$
and $f$ is the fraction of $p(\cdot \vert \theta_j,M_j)$ we want to use.  Often $f$ is taken as a ratio
of sample size to training set size, see \cite{OHagan:1995}.  For suitable chocies of $f$
this BF is said to be good for
impoper priors, non-nested models, and robustness to priors and outliers.
 
Finally, while it is not a BF, it should be remembered that the optimality of BF's is only under generalized zero-one loss
which is insensitive to the degree to which an hypothesis is wrong; it only gives a fixed non-zero value
to incorrect models regardless of how far away they are from a true model.
Thus, for model selection purposes redoing the decision theory for hypothesis testing to
take account of degree of difference would give variants on BF's that had more desireable properties.

\subsubsection{Estimating a predictor}

In addition to testing for model selection to find a predictor, it is possible to estimate a predictor
from the accumulated data.   The simplest is the plug-in predictors:
Given a  parametric family $p(\cdot \vert \theta)$ and good estimator $\hat{\theta}$ of $\theta$
gives an estimated density $p(\cdot \vert \hat{\theta})$ that can be used to generate
point predictors or PI's.   Estimators in plug-in predictors are often frequentist, but may also be Bayes.

The next most familar methods for estimating a predictor are information criteria.
These are mainly for parametric families and the generic form is
\begin{eqnarray}
IC({\cal{D}}, p_{M_j}) = - 2 \log p(y^n \vert \hat{\theta}_j) + \alpha (n)  \dim(\theta_j)
\nonumber
\end{eqnarray}
where $\dim(\theta_j)$ means the number of nonzero elements of $\theta$ in $p_{M_j}$
(i.e., $\| \gamma \|$ in lnear  models) and
$\alpha(n) \rightarrow \infty$ and $\alpha(n)/n \rightarrow 0$ as $m \rightarrow \infty$.
These conditions ensure consistency of model selection.
If $\alpha(n) = \log n$, the familiar Bayes information criteterion (BIC) is obtained.
The Akaike information criterion (AIC) corresponds to taking $\alpha(n) = 2$, but it is not consistent.
IC's are not normally regarded as generating estimators, but they do. Choosing the model that
minimizes an IC, i.e., choosing $\hat{p} = \arg \min IC$, gives a function of ${\cal{D}}$ that identifies a model from
the model list.  It has a sampling distribution which can be used to assess performance.  In the specific case of
the BIC, the minimization is equivlalent to selecting the mode of the across-models posterior;
the mode is the Bayes action under zero-one loss.  Many IC's are consistent and it is reasonable
to use some combination of them for prediction;see \cite{Erven:etal:2012}.

Another general way to estimate a predictor is through regularization.  Let $L_1$ and $L_2$
be loss functions and write
\begin{eqnarray}
\hat{\beta} = \arg \min_{\lambda, \beta}  \left( \sum_{i=1}^n L_1(y_i - f( x_i, \beta )) + \lambda \sum_{i=1}^d L_2=(\beta_j)
\right)
\label{shrinkage1}
\end{eqnarray}
where $f$ denotes a class of models, often linear as in \eqref{lingamma1}.
Optimizating leads to the predictor
$$
\hat{Y}(x_{\sf new}) = f(x_{\sf new}, \hat{\beta})
$$
and this can be generalized to adaptive methods that allow $L_2$ to depend on $j$.
Optima from criteria like \eqref{shrinkage1} often have the oracle property which means that under
sparsity conditions, the
sampling distribution of $\hat{\beta})$ is consistent, asymptotically normal, and efficient.
Importantly, criteria such as \eqref{shrinkage1} correspond to finding the mode
of a posterior obtained from  exponentiating the argument of the objective function,
so $\hat{\beta}$ is a Bayes estimator.


\phantom{
In an observation study, assuming all classical causal assumptions hold, $Y_i (1)$ denotes the potential outcome of person $i$ if they are given the treatment, and $Y_i (0)$ is the potential outcome if they are given the control. We only get to observe one of $Y_i (1)$ or $Y_i (0)$ for each $i$ (the actual outcome), but not both. The average treatment effect ATE $ = 1/n \sum_{i=1}^n (Y_i (1)- Y_i (0))$ is intrinsically a fixed constant.
In light of our distinction between (population) estimation and (individual) prediction, there are two distinct approaches to estimate ATE. From a ``prediction'' point of view, causal inference is all about (counterfactual) prediction---All we need is a generative model of the outcome conditioning on all relevant covariates, and treatment or control, such that we can impute the unobserved $Y_{i}(1)$ for those control units, and  $Y_{i}(0)$ for those treated units. However, a pure ``estimation'' point of view would eliminate the necessity of individual outcome prediction. For example, in methods like matching, if person $i$ is a treatment unit, we will directly match the outcome $Y_i$ with a control person whose covariates are as close to person $i$ as possible; denote this counterpart's index $m(i)$. The difference $Y_i-Y_{m(i)}$ is not meant to provide a good prediction for the individual treatment effect $Y_i (1)- Y_i (0)$; All we ask is the average of this difference is good estimation of ATE when looping over $i$.
Both approaches have pros and cons. The prediction approach well suits the modern common task framework, and could easily facilitate modern machine learning techniques without too much adaptions---except that we need to pay particular attention to out-of-sample prediction since we are making counterfactual predictions---except that we need to care about  out-of-sample predictive performance in any good prediction practice. On the other hand, the estimation approach has not gone away. Its reduced form may deliver a more efficient estimation of ATE because that the individual counterfactual outcomes is redundant if we only want ATE, and should be viewed as nuisance parameters.}

\subsection{Models more generally}

Every valid model must make testable predictions.  Accordingly, every propsoed model class generates
a predictor, possible several.  This includes numerous model classes including
linear models, time series models,  mixed models,  langitudianl models, etc.

The weakenss of using unitary predictors is that they do not incorporate 
model uncertainty:  Unitary predictors from models are conditional on
the model class being true.  That is,  there is a true model and it is in the model 
class -- or at least satisfactorily close to an element in the model class
in a strong enough sense for the application.

There are at least two ways to remedy this.  One is via model expansion in which a proposed model
is situated inside a larger model that is used to assess uncertainty.
Model expansion can be contiuous or discrete.  A second way to address predictor uncertainty is
via model averaging.  This can be regarded as a sort of discrete model expansion.
Continuous model expansion as a way to assess model uncertainty is beyond our present scope
except to note that an expanded model still results in a unitary predictor for comparison purposes.  
However,  the 
next section discusses model averaging.

We note that any way to assign `worth' to a model gives a model average and
model selection technique.
For instance, if $AIC(k)$ is the worth of $M_k$ under the AIC criterion,
then one can choose the model and hence predictor with the highest AIC value.
Alternatively, the $AIC(k)$'s can be normalized to 
$$
AIC^*(k) = AIC(k)/\sum_{\ell=1}^K AIC(\ell)
$$
giving a $K$-vector in the $K-1$ simplex.
The $AIC^*$'s can be used to define a model average and the AIC optimal
model still has the highest $AIC^*$ value.
Thus, model selection and model averaging are interconvertible.

\section{Model Average Predictors}
\label{ModelAverages}

Model averaging has two aspects.  The first is the
average of models on the level of distributions.
The second is a weighted average of predictors from models where the weights 
on the predictors reflect the worth of the models that gave them.   Not all
model averages have both aspects.
Model averages have three main uses, as a single distribution that can be used directly, as a 
comnposite predictor or model, and as a single distribution that can be used
as a `belief' model or predictor for comparison purposes.  Model averages
also typically outperform the components in the averages, see \cite{Clarke:Le:2022},
suggesting that as a rule, composite predictors will outperform unitary predictors,
unless the unitary predictor comes from a model that is sufficiently close to the true model.

The Bayes model average (BMA) is the paradigm example of a model average because
it has essentailly all the desirable properties a model average can have, at least
asymptotically.  Another commonly used model average that is Bayesian is
the median model.  There are other model averages that are frequentist such as
stacking, bagging, and boosting, that often have a Bayesian analog.   
Finally, we discuss some principles for selecting
the models to include in an average.

\subsection{BMA}
\label{BMA}

On the level of densities, the BMA for a future value $Y_{n+1}(x_{n+1})$ is
\begin{eqnarray}
m(y_{n+1} \vert x_{n+1}, {\cal{D}}) = \sum_{k=1}^K W(M_k \vert {\cal{D}}) p(y_{n+1} \vert x_{n+1}, {\cal{D}}, M_k).
\label{BMA1}
\end{eqnarray}
Here,  the posterior weights are
\begin{eqnarray}
W(M_k \vert {\cal{D}}) = \frac{p({\cal{D}} \vert M_k) W(M_k)}{\sum_{i=1}^K p({\cal{D}} \vert M_k) W(M_k) }
\label{BMA2}
\end{eqnarray}
in which
\begin{eqnarray}
p({\cal{D}} \vert M_k) = \int p(y^n \vert x^n,  \theta_k, M_k) w(\theta_k) {\rm d} \theta_k 
\label{BMA2}
\end{eqnarray}
and the posterior predictive density in \eqref{BMA1} is
\begin{eqnarray}
p(y_{n+1} \vert x_{n+1}, {\cal{D}}, M_k) = \int p(y_{n+1} \vert x_{n+1},  \theta_k) w_k(\theta_k \vert {\cal{D}}) {\rm d} \theta_k.
\label{BMA4}
\end{eqnarray}
Expression \eqref{BMA1} is the Bayesian {\it model} average becuse the density on the left
is an average of densities on the right.  It is a model that is the result of averaging other models.

Strictly, the BMA predictor is a predictor derived from the BMA.  Without qualification, the BMA
predictor, often called simply the BMA when no confusion will result, is the Bayes optimal
predicter derived from \eqref{BMA1} under squared error loss.  This is the posterior mean,
\begin{eqnarray}
\hat{Y}_{n+1}(x_{n+1}) &=&  E(Y_{n+1}(x_{n+1}))  \nonumber \\
&=& \sum_{k=1}^K W(M_k \vert {\cal{D}}) E(Y_{n+1}(x_{n+1}) \vert {\cal{D}}, M_k)
\label{BMA5}
\end{eqnarray}
formed using \eqref{BMA4}.  Expression \eqref{BMA5} is the weighted average of predictors from the $K$
models using the posterior model weights.
Under $L^1$ loss, the BMA predictor is the median of the BMA density in \eqref{BMA1};
the median is also considered an average.  Likewise, one could use the mode or any other location
of \eqref{BMA1} to obtain a BMA predictor.

The core idea of a BMA is that the posterior model weights are a reliable indicator of
the worth of a model or predictor so using them to form a weighted sum is going
to be effective.
As noted in Sec. \ref{probclasses}, BMA is optimal in multiple senses.  In addition, 
BMA predictors are optimal in terms of their frequency properties assessed in the 
joint prior distribution of the models and their internal parameters \citep{Madigan:etal:1996}
and the BMA is 
asymptotically optimal from a prequential standpoint,  see \cite{Skouras:Dawid:1998},
but not in general in finite samples, see \cite{Wong:Clarke:2004}.  Moreover, the
latter shows that for prediction it is not in general optimal to condition on all available data.

Despite these good theoretical properties, the BMA has three drawbacks.    First, in general in practice,
its predictive performance often lags other methods.  Second, in
${\cal{M}}$-closed and -complete  cases,  the BMA converges asymptotically to the one single 
model on the list that is closest to the true model in Kullback-Leibler distance, \cite{Berk:1966},
and this can lead to overconfidence, see \citep{Yang:Zhu:2018} and \cite{Huggins:Miller:2023}.
Third, \eqref{BMA4} is sensitive to the prior $p(\theta_k | M_k)$ and \eqref{BMA2} is sensitive to the
prior $W(\cdot)$.


To reduce dependence on the prior,  variants of BMA are sometimes tried.   For instance, using weights in a
BMA based on the LBF, IBF, or FBF is possible.  As an example, $p({\cal{D}}_{\sf min}^c  \vert M_1, {\cal{D}}_{\sf min} )$ from 
\eqref{IBF} can be used in place of the likelihood  in \eqref{BMA2}, with the corresponding changes in
the posterior predictive likelihood in \eqref{BMA1}.

\subsection{Median model}
\label{medmod}

The median predictive model has those explanatory variables that have posterior
inclusion probability at least one-half, see \cite{Barbieri:Berger:2004}.  
That is, suppose $Y = X\beta + \epsilon$ in the usual way
with the $M_k$'s representing all submodels based on different selections of expanatory variables.
Given within-model priors $w_k(\beta_k, \sigma)$, resulting posteriors $w_k(\beta_k, \sigma \vert {\cal{D}})$,
and across models prior $W$, the posterior inclusion probability for a variable $x_j$ is
\begin{eqnarray}
p_j = \sum_{x_j \in M_k} W( M_k \vert {\cal{D}}).
\label{median1}
\end{eqnarray}
It is seen that $W( M_k \vert {\cal{D}})$ is in the sum if and only if $x_j$ is in $M_k$
as an explanatory variable. 
Given $p_1, \ldots, p_K$, the median probability model is the $M_k$ denoted
$M_{\sf med}$ that has exactly those
$x_j$'w with $p_j \geq 1/2$.  Let $l_j = 1$ if $p_j \geq 1/2$,
$l_j = 0$ if $p_j < 1/2$, and  $\ell = (l_1, \ldots , l_J)$; this property justifies the
term median model.   Then,  the $M_{\sf med}$ model is
$Y = x_\ell \beta_\ell + \epsilon$ and the $M_{\sf med}$ model predictor is
\begin{eqnarray}
\label{median2}
\hat{Y}_{\sf med}(x_{n+1} ) = x_{n+1,\ell}  E(\beta_\ell \vert {\cal{D}}),
\end{eqnarray}
where the subscript $\ell$ indicates which explanatory variables in a vector are included.

As noted in \cite{Barbieri:Berger:2004} and \cite{Barbieri:etal:2021}, $M_{\sf med}$ has
some good properties:  It is predictively optimal under $L^2$ and in finite samples outperforms
multiple other predictors in terms of loss and sparsity.  As a rule, these authors observe that
$M_{\sf med}$ is beaten by the BMA (over which it has computational and interpretational advantages) 
but only slightly, an advantage that disappears ans $n$ increases
On the other hand, $M_{\sf med}$ only seems to work
well for certain choices of prior, e.g., $g$-priors,
especially when
the explanatory variables are correlated.  Also, the median predictor effectively requires that
the median model be on the model list, a property called `graphical model structure'.  This is 
not a problem for all subsets regressions but
can be a problem if some models are excluded e.g., in nested settings.  This concept of median model can be extended
beyond linear models but seems not to have been to date.

A variant on the median model predictor is the posterior weighted median (PWM), see \cite{Clarke:etal:2013}.
The motivating observation is that, being a sum, the BMA may be sensitive to terms that are
very large or very small.  So, it may make sense to take a `trimmed mean' form of the BMA or a `median' prediction.  
The idea is to weight each prediction by the posterior probability of the model that
generated it and then take the median of the predictions in the posterior.
To define the PWM formally,
put the predictions in increasing order and then choose one by writing
\begin{eqnarray}
\label{PWM}
\hat{Y}_{\sf PWM}(x_{n+1}) =
 \underset{k = \{1 , \ldots , K\} }{\hbox{med}} [w_k \diamondsuit \hat{Y}_k(x_{n+1} \, | \, \tilde{\beta}_k)] 
\nonumber
\end{eqnarray}
where $\hat{Y}_k$ is the predictor from the $k$-th model, $\tilde{\beta}_k$ is the posterior mean from 
$M_k$ and $\diamondsuit$ is the operation
$$
\underset{k= 1, \ldots , K}{\hbox{med}}[w_k \diamondsuit f_k]  = f_{(r)},
$$
where 
$$
r = \min \{j: \underset{i \leq j, \, i = 1, \ldots , K}{\sum} w_{(i)} \geq 1/2 \}
$$
and $w_{(i)}$ is the corresponding weight of the $i^{th}$ smallest value of the $f_k$'s
over $k = 1, \ldots , K$.   
Often the $w_k$'s are obtained from their usual $BIC_k$ approximations using
MLE's even though the posterior means $\tilde{\beta}_k$ are used to make predictions.  
It is seen that PWM does not require graphical model structure.

The PWM comes from optimizing an $L^1$ criterion using the across-model posterior.  Specifically,
\begin{eqnarray}
\hat{Y}_{PWM}(x_{n+1}) 
&=&  \arg \min_u \sum_{k =1}^K | u - f_k(x^{n+1}  | \tilde{\beta} )| w_k  \nonumber\\
&\approx& \arg \min_u \sum_{k =1}^K | u - f_k(x_{n+1} | \tilde{\beta} )| W(M_k| {\cal{D}})
\label{PWMopt}
\end{eqnarray}
where $W(\cdot | {\cal{D}}_n) $ is the (marginal) posterior distribution for the models in 
${\cal{M}}$.   Note that
in \eqref{PWMopt} the optimum
depends on $x_{n+1}$.  That is, different models may be used to give different 
predictions for different values of $x_{n+1}$ -- unlike other MA methods.
It may be that this increased adaptivity allows the PWM to outperform other predictors
such as the BMA and the median model, if only slightly.

\subsection{Other Model Averages}

We conclude this section with a list of other MA predictors that occur in
practice and often perform well.

\subsubsection{Stacking}

Invented by \cite{Wolpert:1992}, the idea behind stacking is to define coefficients for 
a weighted sum of model-based predictions by
CV.  Given data ${\cal{D}}$ and models $\{ f_k(x \vert \theta_k)$, the stacking pedictor is
\begin{eqnarray}
\hat{Y}_{\sf stack}(x) = \sum_{k=1}^K \hat{\alpha}_k f_k(x \vert \hat{\theta}_k )
\label{stack1}
\end{eqnarray}
where
\begin {eqnarray}
(\hat{\alpha}_1, \ldots \hat{\alpha}_K) = \arg \min_\alpha  \sum_{i=1}^n \left( y_i  - \sum_{k=1}^K \alpha_k  f(x_i \vert \hat{\theta}_{k, \righthalfcap i}) \right)^2
\label{stack2}
\end{eqnarray}
and $(y_i, x_i)$ is not used to estimate $\hat{\theta}_{k, \righthalfcap i}$.  Leave $m$-out CV can be used
in \eqref{stack2} in place of leave-one-out CV and $\hat{\theta}_{k, \righthalfcap i}$ can be used in place 
of $\hat{\theta}_k$ in \eqref{stack1}.   Often, the posterior mean in $M_k$ is used to estimate $\theta_k$.
The empirical quantity minimized in \eqref{stack1} corresponds to the population quantity
\begin{eqnarray}
\int  \left( f(x) - \sum_{k=1}^K \alpha_k f_k(x) \right)^2 {\rm d} \mu(x)
\nonumber
\label{stack3}
\end{eqnarray}
when $f$ is the true model and has dominating measure $\mu$.

The stacking procedure can be applied to model densities, see \cite{Smyth:Wolpert:1998}, predictive densities
see \cite{Yao:etal:2018}, and many other quantities such as posterior densities.  These latter
methods would enable intervals for prediction (or estimation).   The role of CV itself
leads to an action $\hat{\alpha}$ that is asymptotically Bayes optimal under several
loss functions including squared error and log, see \cite{Le:Clarke:2017}.
Stacking can also be generalized from squared error in \eqref{stack2} to any loss function or score function,
\cite{Yao:etal:2018}.

Ostensibly for the ${\cal{M}}$-closed case,  stacking was brought into the ${\cal{M}}$-complete and -open settings by
\cite{Clyde:Iversen:2013} who also observed that in those settings relaxing stacking weights to merely requiring positivity
(but not sum-to-one) might improve predictive performance.  \cite{Breiman:1996} made the observation that
the more different the models in the `stack' were, the better performance could be expected.  This was corroborated in
\cite{Clarke:2003} who also gave conditions ensuring stacking weights would be consistent
and showed that stacking could be more robust than BMA, possibly owing to the use
of CV to find the $\hat{\alpha}_k$'s.  

\subsubsection{Bagging}

Bagging is a portmanteau of `bootstrap aggregating' which is essentially the
idea of getting a collection, usually $n$, of bootstrap samples, using each sample to obtain
a data-dependent quantity, and then somehow combining the $n$ quantities.  In prediction,
bagging is often used on point predictors:
\begin{eqnarray}
\label{bag1}
\hat{Y}_{\sf boot} (x_{n+1}) = \frac{1}{B} \sum_{b=1}^B \hat{Y}(x_{n+1} \vert {\cal{D}}_b)
\end{eqnarray}
where $B$ is the number of bootstrap samples $ {\cal{D}}_b$ drawn from {\cal{D}}.
Bagging is typically applied to good but unstable predictors as a way to stabilize them.
Moreover, as a result bagged point predictors are nearly Bayesian, see \cite{Clarke:Le:2022}.

Bagging the across-models posterior gives
\begin{eqnarray}
W_{\sf boot}(k) = \frac{1}{B} \sum_{b=1}^B W(k \vert {\cal{D}}_b).
\label{bag2}
\end{eqnarray}
This is studied in \cite{Huggins:Miller:2023} who argue that, as expected, 
using the bagged posterior for model selection provides stability.  This is
particularly important when models are
so close together it's hard to distinguish them from each other.  In such cases,
the bagged posterior converges asymptotically to a {\it Bernoulli(1/2)} -- 
an intuitively reasonable property.
In principle,  the techniques of \cite{Huggins:Miller:2023}
can be applied to posterior predictive distributions and results
analogous to their Theorems 3.1 and 3.2 and  Corollary 3.3 obtained.
Indeed, any posterior quantity, including BF's for instance is amenable to bagging.

\subsection{Notes to Secs.  \ref{Predsel} and \ref{ModelAverages}}

We end this section with a
recitation of methods that would have been discussed had space allowed.

First, there are many unitary classifiers (predictors of categorical $Y$'s) that are
in regular use.    One is the naive Bayes classifier that amounts to maximizing the posterior
(predictive) probability given $x$ under the assumption that the $x_j$'s are independent. 
That is, the predictor for $Y_{n+1}(x_{n+1})$ is
\begin{eqnarray}
\hat{Y}_{\sf NB}(x_{n+1}) = \arg \max_\ell P( Y(x_{n+1}) = \ell ~\vert x_{n+1}, {\cal{D}})
\label{Notes45_1}
\end{eqnarray}
where $\ell = 1, \ldots, L$ are the class values $Y$ can take.  Expression \eqref{Notes45_1}
is parallel to \eqref{argminpostrisk} in that $\hat{Y}_{\sf NB}$ is Bayes optimal.  This means
that if the risk of an action $\delta$ is
$R(\delta) = P(\delta \neq Y)$ then, $R(\hat{Y}_{\sf NB}) \leq R(\delta)$ for any delta.
More explicitly,  for $L=2$, 
\begin{eqnarray}
\hat{Y}_{NB}(x) = \begin{cases} 1 & P(Y(x) =1 \vert x) \geq 1/2 \\
1 & P(Y(x) =1 \vert x) < 1/2 \\
\end{cases}
\label{Notes45_2}
\end{eqnarray}
and $\hat{Y}_{NB}$ may have parameters or a density that must be estimated using
${\cal{D}}$, e.g., by the MLE or Nadaraya-Watson.   The independence assumption simplifies the classifier enormously
especially for large $J$ but may not in fact lead to poor performance, see \cite{Hand:Yu:2001}.
There are as many Bayes classifiers as there are $P$'s, the most typical being normal.

The logistic classifier is defined by setting
\begin{eqnarray}
P(Y=1 \vert x) = \frac{1}{1 + e^{\alpha_0 + \sum_{j=1}^J \alpha_j x_j}}
\label{Notes45_3}
\end{eqnarray}
with $P(Y=0 \vert x) = 1 - P(Y=1 \vert x)$ when $L=2$.  The rule is to set $\hat{Y}_{\sf Log}(x)$ to be
zero if and only if $P(Y=0 \vert x)/P(Y=1 \vert x) > 1$ which is equivalent to
\begin{eqnarray}
\alpha_0 + \sum_{j=1}^J \alpha_j x_j > 0.
\label{Notes45_4}
\end{eqnarray}
Bayes or frequentist estimatoers can be used for the $\alpha_j$'s, but MLE's are the most common.

There are numerous other unitary classifiers including Fisher's discriminant, support vector machines,
relevance vector machines, etc. and each conditions under which it is optimal or at least performs very well.

By contrast, boosting is a composite classifier.  It's traditional form is frequentist.  The boosting
classifier is a sum of iterates where each new iterate is formed by adjusting the loss structure
to make errors more costly.    The iterates are often called weak learners even if this is a misnomer.
Any classifier can be boosted -- included the naive Bayes classifier,
but it is unclear how much improvement to expect.  Bayesian boosting has also been proposed,
see \cite{Lorbert:etal:2012}.  This is different from gradient boosting which is also a 
model average but based on gradient descent not changing the loss function.
There are efforts to apply the intuition for boosting to regression problems, but it is unclear
how much improvement this gives in general.

Three final promising methods: 1) Bayesian additive regression trees.  This is a Bayesian MA
for regression based on using the stumps or trees as if they were a sort of basis expansion,
see \cite{Hill:etal:2020} for a recent review.
2) Log-linear pooling.   Proposed as a way to pool prior information, see \cite{Rufo:etal:2012},
 the idea is to write
$w(\theta) \propto \prod_{k=1}^K w_k(\theta)^{\alpha_k}$ and choose the $\alpha_k$'s from the $K-1$ dimensional
simplex using a loss function.  However, the same formula can be used to
combine posteriors or posterior predictives.  3) Bayesian predictive synthesis.  The idea is to
obtain predictions from
\begin{eqnarray}
p(y \vert {\cal{H}}) = \int \alpha(y \vert x) \prod_{k=1}^K h_k(x_k) {\rm d}x
\label{Notes45_5}
\end{eqnarray}
where the ${\cal{H}} = \{ h_1, \ldots, h_K\}$ are forecast distributions from $K$ experts, 
$x$ is a latent variable, and $\alpha(y\vert x)$ is a given conditional density.  There may be a class of viable densities 
satisfying \eqref{Notes45_5}, see \cite{McAlinn:West:2022} for details.


\section{Streaming data}
\label{streaming}

Abstractly, streaming data means that we receive one data point at a time and seek to predict the next
data point, or perhaps next $k$ data points, using the accumulated data.   We might also be interested
in making a decision at each time point; this is a generaliztion of the prediction problem.
However,
data may arrive irregularly and our predictors may have both a storage and running time constraint.
That is,  we often require the predictor for time $n+1$ to be easily updatable from the predictor from time $n$.
In addition, streaming data is rarely ${\cal{M}}$-closed.  Usually they are ${\cal{M}}$-complete or -open.
In the former case,  predictive methods from time series are often relevant.  
In the latter case, time series or other stochastic methods may be invoked, but the absence
of a probabilistic structure for the data itself often means we must include other classes of methods.
So, here we present
two approaches that are specifically designed for the general streaming case.
We begin by recalling some basics of Bayesian time series modeling and then turn to
a Bayesian Shtarkov predictor for strings of data and to relatively recent methods based on `hash' functions
for streaming data.   

\subsection{Time Series}

Very loosely,  Bayesian predictions from time series methods stem from one of two approaches:  Box-Jenkins models
and state space models.  Either can be multivariate but here we limit attention to
univariate cases.

\subsubsection{Bayesian Box-Jenkins}
\label{BBJ}

In the simplest formulation of the streaming data problem we have a stochastic process $Y_1, Y_2, \ldots$
in which relationships among the $Y_i$'s are given in terms of the backshift operator $B$ that acts on
the individual random variables to give their one step earlier version, that is, $B(Y_{n+1}) = Y_n$.
Often polynomials in $B$ are used such as $\phi(B) = 1 - \phi_1 B - \cdots - \phi_p B^p$ where $p$
is called the order of $\phi$ and for convenience we write $\phi =(\phi_1, \ldots, \phi_p)$ for the parameters.  
Now an auto-regressive moving
average model is written as
\begin{eqnarray}
\phi(B) Y_n = \theta(B) \epsilon_n
\label{ARMApq}
\end{eqnarray}
for each $n$
in which $\theta$ is a polynomial of order $q$ in $B$ analogous to $\phi$ and the $\epsilon_n$'
are IID $N(0, \sigma^2)$.  It is straightforward to put a prior on $\phi$, $\theta$, and $\sigma$:
Using \eqref{ARMApq},  write $\epsilon_n$ as the difference between $Y_n$ and polynomials $\phi$ and $\theta$ 
in $B$ acting on $Y_n$:
\begin{eqnarray}
\epsilon_n = Y_n - \sum_{j=1}^p \phi_j Y_{n-j} - \sum_{j=1}^q \theta_j \epsilon_{n-j}.
\label{isolateeps}
\end{eqnarray}
Since the $\epsilon_i$'s are IID,  using \eqref{isolateeps} the likelihood is
\begin{eqnarray}
L( \phi, \theta, \sigma \vert Y^n = y^n) \propto \left( \frac{1}{\sigma^2}\right)^{(n-p)/2} e^{\frac{1}{2\sigma^2} \sum_{t=p+1}^n \epsilon_t^2}.
\nonumber
\end{eqnarray}
Equipping $\phi$, $\theta$, and $\sigma$ with priors, and assumning an initial distribution
for $Y_0$ (that we have ignored here for simplicity) leads to the posterior
$w(\phi, \theta, \sigma \mid y^n)$ and therefore to a posterior predictive $p(y_{n+1} \vert y^n)$,
see \eqref{postpred1}.   This process can be used to give a predictive density for multiple $Y_i$'s and for
more complicated models that introduce seasonality and `integration' -- a differencing of
the time series separated by say $d$ time points.   The effect of these is to make the polynomials
in \eqref{ARMApq} more complicated.  Point predictors for $\hat{Y}_{n+1}$ can be derived
from $p(y_{n+1} \vert y^n)$ by taking the posterior mean, for instance, and give PI's by
using the posterior variance.   These are computational issues that we do not address here as they have been
amply addressed elsewhere.

\subsubsection{Bayesian State Space Models}
\label{BSS}

A state space model (SSM) is a generalization of a Box-Jenkins model that represents the observation at time $i$,
i.e.,  $Y_i$,  as a function of an underlying state $X_i$ with $\dim(X_i) = d$ that satisfies a transition equation.
With initial state $X_0$,  the sequence of random variables is 
\begin{eqnarray}
    X_0 \longrightarrow  \begin{pmatrix}
           X_{1} \\
           Y_{1}
         \end{pmatrix}
\rightarrow
\begin{pmatrix}
           X_2 \\
           Y_2
         \end{pmatrix}
\longrightarrow \cdots
\begin{pmatrix}
           X_{n+1} \\
           Y_{n+1}
         \end{pmatrix} 
\label{HMM}
\end{eqnarray}
and it is implicitly
assumed that $Y_i$ is available after $X_i$ and $X_{i+1}$ is available
after $Y_i$.
Typically, it is also assumed the $X_i$'s are Markov and the 
`transition' equation can be written $X_i = f(X_{i-1}, \eta_i)$ for some $f$, 
where the $\eta_i$'s are IID noise.  Also,  the `observation' equation can be written
$Y_i = g(X_i, \epsilon_i)$ for some $g$, where the $\epsilon_i$'s are also IID.   Thus, an SSM is, at root,
a hidden Markov process. 

One reason \eqref{HMM} is called an SSM is that one of the main tasks
is to identify the underlying `state' $X_i=x_i$ from the observations $y_1^n$.   When $f$ and $g$ are linear and the noise terms $\eta_i$ and 
$\epsilon_i$ are normal,  the set of updating equations (mean and variance)
for $X_{n+1}$ given $Y^n$ is often called Kalman filter prediction; the  corresponding set of
expressions for $X_{n+1}$ given
$Y^{n+1}$ is often called Kalman filter updating.  The term `filter' refers to the
fact that as $n$ increases, the conditioning $\sigma$-fields 
$\sigma(Y^n)$ or $\sigma(Y^{n+1})$ increase.   The analysis of \eqref{HMM} is `Bayesian' in
that Kalman filters for prediction or updating look like posteriors for $X_{n+1}$ given past observations
$Y^n= y^n$ or $Y^{n+1} = y^{n+1}$.    

To be more precise,  we follow the derivation in \cite{Gurajala:etal:2021} for Kalman filter prediction.
For a more detailed and equally lucid treatment, see \cite{Petris:etal:2008} Chap.2.  
Let $\eta_i \sim N(0, Q)$ and $\epsilon_i \sim N(0, R)$ and
fix  matrices $F$, $G$ and $H$ so that
\begin{eqnarray}
\nonumber X_{n+1} &=& F X_n + G \eta_n \\ \nonumber
Y_{n+1} &=& H X_n + \epsilon_n.
\end{eqnarray}
This structure is often called dynamic linear regression, especially when $F=F_n$, $G=G_n$ and $H= H_n$.
Since all variables are normal and all transfomations are linear,  it is easy to derive
a point predictor for $(X_{n+1} \vert Y^n= y^n)$ and its variance as
\begin{eqnarray}
\nonumber
\hat{X}_{n+1} &=& F E(X_{n+1} \vert Y^n= y^n) \\ \nonumber
\hbox{Var}(X_{n+1} \vert Y^n = y^n) &=& F \hbox{Var}(X_n \vert Y^n = y^n) F^\top + GQG .
\end{eqnarray}
Similar expressions hold for Kalman filter updating.  The standard assessment of Kalman filter convergence
is how well the PI's formed from mean and covariance match the data over time.  When the various
parameters are accurate,
$\hat{X}_{n+1}$ becomes a linear function
and $\hbox{Var}(X_{n+1} \vert Y^n = y^n)$ goes to a constant at rate ${\cal{O}}(1/n)$.

Given the expressions for Kalman filter prediction, $Y_{n+1}$ can be predicted from
\begin{eqnarray}
p(y_{n+1} \vert y^n) = \int p(y_{n+1} \vert x_{n+1}) p(x_{n+1} \vert y^n) {\rm d} x_{n+1},
\label{obspred}
\nonumber
\end{eqnarray}
parallel to \eqref{postpred1}.

In a fully Bayesian treatment $F$, $G$ , $H$, $Q$, and $R$ would be regarded as 
parameters equipped with a prior.    Such a treatment would lead to a more flexible model with intuitive inferences. 
Bayesian models also allow specification of non-Gaussian state disturbances and observation innovations.
However, this more general treatment does not seem to have been carried out.

\subsection{Shtarkov solution}
\label{Shtarkov}

The Shtarkov solution, see \cite{Shtarkov:1987}, is a predictor for outcomes $y_{n+1}$ 
given $y^n$ where all $y_i$'s come from an ${\cal{M}}$-open DG; sometimes this is called prediction along a `string'. 
This sequential prediction problem is a game between Nature $N$ and a Forecaster $F$
in which, at each round, $F$ tries to predict the $y_{n+1}$ that $N$ will produce and $N$ is then
free to choose $y_{n+1}$ by any rule s/he wants -- or by no rule at all.  

It is assumed that $F$ has access to a collection of `experts' i.e., predictions from rules, indexed by $\theta$.
The experts are denoted $p(\cdot \vert \theta)$ and we assume a `prior' $w(\cdot)$ over $\theta$.
Strictly, $w$ is only a weighting function on $\theta$ and $w(\theta)$ represents
our pre-data view of the reliability of expert $\theta$.  That is, $p_\theta$ is regarded as an
action that $F$ might use.

To define the game, we must specify a utility function.  In the Shtarkov solution we use $\log$
because of its role in source coding.  For instance, the Shannon code
has codelengths of the form $\lceil \log P(Y=y) \rceil$ for discrete $Y$.
Now, the $n+1$ round of the game proceeds
as follows. Each expert, indexed by $\theta$, announces a density $p_\theta$.
Given this,  $F$ announces a density $q(\cdot)$ that will be used to predict the
value $N$ issues. Finally, $N$ issues $\hat{y}_{n+1}$ and pays $F$ $\log q(y)$. If this number is negative, 
it is the amount of money $F$ pays $N$,
concluding the round.    

The Shtarkov solution is what $F$ chooses for $q$ by minimizing the maximum regret.
The regret is the extra cost $F$ incurred over how the best expert would have performed.
That is, the point of the Shtarkov solution is to mimic the performance of the best expert.
Formally, we want to use 
\begin{eqnarray}
q_{\sf opt}(y_{n+1} \vert y^n) =  \frac{q_{\sf opt}(y^{n+1})} {q_{\sf opt}(y^{n})}  
\nonumber
\end{eqnarray}
where for any $n$, $q_{\sf opt}(y^{n})$ is
\begin{eqnarray}
 \arg \inf_p \left[ \sup_{y^n} \left( \log \frac{1}{q(y^n)} -  \inf_\theta \log \frac{1}{w(\theta)p(y^n \vert \theta)}             \right)           \right].
\label{Shtarkov1}
\end{eqnarray}
The difference in expression \eqref{Shtarkov1} represents how much worse $F$'s action $q$ is than
the best expert i.e., $\theta$, is.
It is seen that
\begin{eqnarray}
q_{\sf opt}(y^{n}) &=& 
\arg \inf_p \left[ \sup_{y^n} \sup_\theta \left(  \log \frac{w(\theta)p(y^n \vert \theta)}{q(y^n)}  \right)       \right]
\nonumber \\
&=&  \frac{w(\tilde{\theta})p(y^n \vert \tilde{\theta})}{ \int w(\tilde{\theta})p(y^n \vert \tilde{\theta}){\rm d} y^n },
\nonumber
\end{eqnarray}
where $\tilde{\theta}$ is the posterior mean;
see \cite{Shtarkov:1987}  for further details.  Replacing the
integral with a sum gives the expression for discrete $y^n$.  For computational details, see
\cite{K:M:2007} and \cite{Le:Clarke:2016}.

The behavior of the Shtarkov solution for ${\cal{M}}$-closed and -complete problems
can be determined, but its usefulness is primarily with observational data that is often
${\cal{M}}$-open.  Indeed, the fastest growing class of data may be large volume
observational data collected by automated methods such as web-scraping.
The limitation of this method is that it has mainly been implemented only for discrete $y$'s.
However, it can be extended to continuous $y$'s and equally important to nonparametric
priors in place of $w$.

\subsection{Hash Functions}
\label{hash}

In the 80's, computer scientists recognized that there may be no reasonable
probabilistic structure for ${\cal{M}}$-open streaming data.   
This is important because
one of the fastest growing data types -- that does not have a name yet  -- results from automated collection
of large volumes of observational data.   
So, they developed a way to put a probabilistic structure on the objects used for inference rather
than the data.  In this sense, they are being Bayesian.     Given that probabilistic structure,
one can form a streaming analog of a mean, median, or other point predictor or in fact of
a whole distribution function as well as many other statistical quantities such as range,
see  \cite{Muthu:2009} for numerous examples.
Such algorithms are called streaming if they process the data stream so as to provide estimates in one pass
subject to a memory constraint in terms of the sample size, e.g., $\log n$.  Most of these streaming algorithms
are sketches, meaning they compress the data stream so that some function of the stream can
be effectively computed.  Sketches are assumed combinable in the sense that given
two data streams there is a way to derive the sketch for the concatenated stream from the individual streams.

Let $[U] = \{ 1, \ldots, U\}$, with $[V]$
defined similarly, and 
\begin{eqnarray}
{\cal{H}} \subseteq \{ h: [U] \rightarrow [V] \}.
\end{eqnarray}
The class ${\cal{H}}$ is called a hash family and the elements of ${\cal{H}}$ are
called hash functions.   Clearly $\mathbf{card} ({\cal{H}}) \leq U^V$ and it would take at most $V \log U$
bits to encode all of ${\cal{H}}$.  If ${\cal{H}}$ is too large, then $U$ or $V$ is too large and coding is infeasible.  
So, assume a probability distribution $W$ on ${\cal{H}}$ with the property that $\forall u, u^\prime ~ \forall v, v^\prime$:
\begin{eqnarray}
 u \neq u^\prime \implies 
W( \{H(u) = v\} \cap \{H(u^\prime) = v^\prime \}) = \frac{1}{V^2},
\label{2universal}
\end{eqnarray}
where $H$ is the random variable with outcomes $h \in {\cal{H}}$.
The property \eqref{2universal} 
is called 2-universal 
because it applies to pairs of elements drawn from all
of ${\cal{H}}$.
Usually, $V \ll U$ so it will be typical for $h$ not to be one-to-one.  
A `collision' occurs for $h$ when there are $u, u^\prime \in [U]$
so that $h(u) = h(u^\prime)$.    
In fact, under \eqref{2universal},  this happens with probability $W(H(u) = H(u^\prime)) = 1/V$.
Clearly, the smaller $V$ is, the more fewer collisions will occur, and the greater
the data compression will be.
The goal in choosing ${\cal{H}}$ is to give up some of the one-to-oneness of the $h$'s
to allow the functions to be coded with less storage by
allowing for a smaller $V$.  To get around the resulting collisions, one chooses
multiple hash functions and combines them, the storage for multiple hash functions
being less than the storage for the single correct function.

The suprising point is this:   Using hash functions, it is possible to give a sketch of a 
streaming algorithm that predicts the next outcome.  

One example of this is
the {\sf Count-Min} sketch, for data streams $y_1, y_2, \ldots$
that assume values in a finite set of known (large) size that 
we take to be $[U]$, see \cite{Cormode:Muthu:2005}. 
Write $a(i) = (a_1(i), \ldots  , a_U(i))$ where 
$$
a_u(i) =  \mathbf{card} \left(  \{ y_j \vert ~  j \leq i ~  \mbox{and}  ~ y_j = u\} \right),
$$
for $u \in [U]$.  That is, the vector $a(i)$ is the number of occurrences of each  $u=1, \ldots , U$ up to time $i$.
We update $a(i)$ to $a(i+1)$ upon receipt of $y_{i+1}$ by incrementing the
$u$-th element of $a(i)$ by one.  That is, 
\begin{eqnarray}
a_{u}(i+1) = \begin{cases} 
 a_{u}(i) + 1  &  \mbox{if} ~ y_{i+1} = u \\
a_{u}(i) &  \mbox{if} ~  y_{i+1} \neq u 
\end{cases}
\label{stream}
\end{eqnarray}
Obviously, $(1/n) a(n)$ is a probability vector on $[U]$ at time $n$.  So,
in a limiting sense, $a(n)/n$ gives a distribution function $\hat{F}_n$ on $[U]$.
Now, $\hat{F}_n$ can be used to generate a prediction for $y_{n+1}$ by
taking a mean, median, or other location estimator and choosing the 
the value of $[U]$ closest to it as a point predictor.  This can easily be extended
to continuous streams and PI's, but we do not do this here.
However, there are at least three problems with this approach if $U$ is very large.   First, it can be inefficient
if we want to look at the data in one pass.  Second, we want to control the storage.
Third, we want to control
the error even if it is only in a streaming sense.
We can achieve all three by using hash functions.     

The {\sf Count} part of the {\sf Count-Min} sketch is the following.
Let $\epsilon, \delta >0$.  Choose $d = \lceil \log (1/\delta) \rceil$ hash functions
$h_1, \ldots,h_j, \ldots h_d$ at random from ${\cal{H}}$ equipped with a probability that satisfies \eqref{2universal}
and, assuming $U$ is very large,  set $W = \lceil 2/\epsilon \rceil$, where $W \ll U$ is the degree of compression.
For each $i$, form a $d \times W$ matrix $C(i) = (c_{jk}(i))$ that is 0 for $i=0$ and for each $i \geq 1$ is
$c_{jw}(i) = {\sf Count}(j, h_j(y_i))$ where the function ${\sf count}$ updates as $a_u$ does
in \eqref{stream} but using
$W$ in place of $U$ thereby allowing collisions.  That is, the $(j,w)$ element of $C(i)$ is the the number of times the $j$-th
hash function has assumed the value $w \in [W]$ on the elements of the sequence $y_1, \ldots, y_i$.

The {\sf Min} part of the {\sf Count-Min} sketch is the following.  Given the count matrix $C(i)$,
we choose the minimum entry over the $d$
elements in each of the $w$ columns. and write this as $\hat{F}_n(w) = \min_{j=1}^d c_{jw}(n)/n$.
Now for each $w$ we have an empirical probability that $y_{n+1}$ will be $w$.  The resulting 
DF $\hat{F}_n(\cdot)$ is
empirical in that it gives a probability vector for the set $[W]$ computed from data, however, there is no guarantee that
$\hat{F}_n$ converges.  

The output of the {\sf Count-Min} sketch, $\hat{F}_n$, has the nice properties we want.  Specifically, it is readily
computed iteratively i.e., in one pass, and it is combinable in the sense that if we have two streams we can 
add the minima of their respective $\hat{F}_n$'s.  Less obviously,  we can control the error and the
storage required.    For the error statement,  write $F(w)$ to mean the actual proportion of $y_i$'s in the stream
assuming value $w$.  Then,
\begin{eqnarray}
F(w) \leq \hat{F}_n(w) \leq F(w) + \epsilon.
\label{streamingbounds}
\end{eqnarray}
The first inequality in \eqref{streamingbounds} holds by construction and the second inequality holds
with probability at least $1-\delta$ where the probability is on the hash functions in $\hat{F}_n$, see \cite{Muthu:2009}
for the details of proof.
For the storage bound, 
it can be shown that the space requirement for the {\sf Count-Min} sketch is bounded by $(1/\epsilon)\log(1/\delta)$,
see Cor. 1 in  \cite{Cormode:Muthu:2005} or Sec. 5 in \cite{Chakraborty:2020}.

Importantly,  the {\sf Count-Min} sketch is only one of many sketches based on
hash functions for use with streaming data to predict a future outcome or make other decisions sequentially
-- and prequentially.

\section{Concluding Remarks}
\label{conclusion}

This paper rests on what might be called Berger's dictum: Anything that works well is Bayes or 
almost Bayes (although it is not always easy to show that is so for everything that works well).
This is the heuristic behind the complete class theorem as noted in the Editorial.  Thus,
Bayes prediction can serve as a proxy for all good prediction provided three criteria are met:
1)  The predictand and the predictor must be well-defined; 2)  There has to be concept of 
sequential improvement in prediction or at least a way to assess the worth of predictors; and,
3) There has to be something stable enough  enough to be predicted.
We note that prediction is more demanding than modeling because a predictor can be readily
falsified.  Indeed, a model is often little more than summarization of the data whereas a 
predictor goes beyond data summarization to give something testable with future outcomes.

Not less than for distinct approaches to Bayesian prediction have been named, organized, reviewed; the stance of this
paper is neutral toward them.  The key contribution here is to structure the ideas and parlance of
Bayesian prediction, and prediction more generally.  Hence, our distinction between
unitary and composite predictors,  and between algorithmic and model-based predictors.
Also, our effort to be complete within the context of providing a limited space
presdentation of the core ideas.  

Nevertheless,  the overview presented here is incomplete.  Indeed, in a field as creative and 
rapidly evolving as Bayesian prediction, completeness is at best a siren song.
No work can be encyclopedic except an encylopedia.

We conclude with a (partial and biased) list of questions that we think are the most pressing open problems
for Bayesian prediction.

\begin{itemize}

\item How can we combine 
the information'in multiple data types with different properties into one predictor?
Some important contributions have been made but the general line of inquiry invites new methods.

\item How do we assess the real world correlates of the components in a unitary predictor?
Thisn includes consistency -- is the component representative of the correlate?  Do we have
a useful level robustness of the components?  An example:  Do the subnets of a neural
net have any physical meaning?

\item What models should be on the model list for predictor selection or model averaging?
There is a variance-bias tradeoff on the level of model lists that should be optimized.
This follows from the observation that smaller model
lists may give bias and larger model lists may increase 
(posterior) variance.   Dilution priors are one way to address this partially but further heustics for
model list selection should be developed.

\item How do we effectively search general classes of predictors to find a good one and then
automate the reduction of a good predictor to a viable model?  This ensures that any m odel
proposed will have quantified predictive properties and hopefully alleviate poor fit and problems
with reproducibility.

\item How do we assess predictor/model uncertainty and mis-specification relative to general
classes of predictors/models, not just to the list of predictors/models we have used.

\end{itemize}

This is an invitation, particularly to junior people, to contribute their novel insights to one of the
most exciting and hopefully influential field in contemporary Statistics.  The contributions may be
of all sorts -- applications, theory, computing, and, perhaps most important, as this paper has sought to
emphasize, philosophy.   Indeed, scratch a Bayesian, find a philosopher.


\begin{thebibliography}{4}


\bibitem{Abdellaoui:Wakker:2020}
\textsc{Abdellaoui, M. and Wakker, P.}(2020) Savage for dummies and experts. 
{\it J. Econ. Theory.} {\bf 186}, 1-20.


\bibitem{Aitchison:1975}
\textsc{Aitchison, J.}(1975)  Goodness of prediction fit.
{\it biometrika}, {\bf 62},  547-5



\bibitem{Aitchison:Dunsmore:1975}
\textsc{Aitchison, J. and I. R. Dunsmore} (1975)
{\it Statistical Prediction Analysis}, Cambridge University Press, Cambridge UK.

\bibitem{Alcantara:etal:2023}
\textsc{Alcantara, I., Naranjo, J., and Lang, Y.}(2023)
Model selection using PRESS statistic. {\it Comp. Stat.}, {\bf 38},285–298.

\bibitem{Bahri:etal:2018}
\textsc{Bahri, M. Maniu, S. and Bifet, A.}(2018) Sketch-based naive {B}ayes algorithms for evolving data streams.
{\it IEEE International Conference on Big Data}

\bibitem{Barbieri:Berger:2004}
\textsc{Barbieri, M. and Berger, J.}(2004)
Optimal predictive model selection. {\it Ann. Stat.}, {\bf 32}, 870-897.

\bibitem{Barbieri:etal:2021}
\textsc{Barbieri, M., Berger, J., George, E., Ro\v{c}kov\'{a},V.}(2021)
The median probability model and correlated variables.  {\it Bayesian Anal.}, {\bf 16}, 1085-1112.

\bibitem{Bates:Granger:1969}
\textsc{Bates, J. M. and Granger C. W. } (1969) The combination
of forecasts.  {\it Oper. Res. Quart.} {\bf 20}, 451–468.

\bibitem{Berger:Pericchi:1996}
\textsc{Berger, J. O., Pericchi, L.}(1996)
The Intrinsic Bayes Factor for Model Selection and Prediction.  {\it J. Amer. Stat. Assoc. },
{\bf 91}, 109-122.

\bibitem{Berger:Pericchi:2001}
\textsc{Berger, J. O., Pericchi, L.}(2001)
Objective Bayes Methods for Model election:  Introduction and Comparison.
In:  P. Lahiri, Editor, {\it Model Selection}, IMS Lecture Notes Monograph Series Vol. 38,  135-207.

\bibitem{Berk:1966}
\textsc{Berk, R.}(1966) Limiting Behavior of Posterior Distributions when the Model is Incorrect.
{\it Ann. Math. Statist. }, {\bf 37}, 51-58.

\bibitem{Bernardo:Smith:2004}
\textsc{Bernardo, J. M. and Smith, A. F. M. } (2004). \textit{Bayesian Theory}
Wiley Series in Probability and 
Statistics.  Chichester, UK.

\bibitem{Bersson:Hoff:2022} 
\textsc{Bersson, E. and Hoff, P.}(2022) Optimal conformal prediction for small areas.
arXiv:2204.08122, \url{https://doi.org/10.48550/arXiv.2204.08122}

\bibitem{Berti:etal:2021}
\textsc{Berti, P., Dreassi, E., Pratelli, L., and Rigo, P.}(2022) A class of models for Bayesian
predictive inference. {\it Bernoulli}, {\bf 27}, 702-726.

\bibitem{Berti:etal:2022}
\textsc{Berti, P., Dreassi, E., Leisen, F., Pratelli, L., and Rigo, P.}(2022)
A predictive approach to Bayesian forecasting. \url{https://arxiv.org/pdf/2208.06785.pdf}

\bibitem{Brehmer:Strokorb:2019}
\textsc{Brehmer, J. and Strokob, K.}(2019) Why scoring functions cannot assess tail probabilities.
{\it Elec. J. Stat.}, {\bf 13}, 4015-4034.

\bibitem{Breiman:1996}
\textsc{Breiman, L. }(1996). 
Stacked regressions.  {\it Machine Learning,} {\bf 24},49–64.



\bibitem{Chakraborty:2020}
\textsc{Chakraborty, A.}(2020)
{\it Data Stream Algorithms},  Lecture notes, see \url{https://www.cs.dartmouth.edu/~ac/Teach/data-streams-lecnotes.pdf}

\bibitem{Clarke:2003}
\textsc{Clarke, B.}(2003)
Comparing Bayes model averaging and stacking when model approximation error
cannot be ignored.  {\it J. of Mach. Learning Res.}, {\bf  4},683–712.


\bibitem{Clarke:2007}
\textsc{Clarke, B.}(2007) Information optimality and Bayesian modelling.  {\it J. Econometrics}, {\bf 138}, 
405-429.

\bibitem{Clarke:Clarke:2009}
\textsc{Clarke,  J. and Clarke, B.}(2009)
Prequential Analysis of Complex Data with Adaptive Model Reselection. {\it Stat. Anal.  Data Min.},
{\bf 2}, 274-290.

\bibitem{Clarke:etal:2013}
\textsc{Clarke, J., Clarke, B., and Yu, C.-Y.}(2013)
Prediction in {\cal{M}}-complete problems with limited sample size.
{\it Bayesian Analysis}, {\bf 8}, 647-690.

\bibitem{Clarke:Clarke:2018}
\textsc{Clarke, B. and Clarke, J.}(2018) {\it Predictive Statistics:  Analysis and Inference Beyond Models}.
Cambridge University Press, Cambridge, UK.

\bibitem{Clarke:Le:2022}
\textsc{Clarke, B. and Le, T.}(2022)
Model Averaging Is Asymptotically Better Than Model Selection For Prediction.
{\it JMLR}, {\bf 23}, 1-53.


\bibitem{Clemen:1989}
\textsc{Clemen,  R. }(1989) Combining Forecasts:  A review and annotated bibliography.
{\it Int. J. Forecasting}. {\bf 5}, 559-583.

\bibitem{Clyde:Iversen:2013}
\textsc{Clyde, M. and Iversen, E. }(2013)
Bayesian model averaging in the {\cal{M}}-open framework. In:
{\it Bayesian Theory and Applications}, 483–498. Oxford University Press.

\bibitem{Cormode:Muthu:2005}
\textsc{Cormode, G.  and Muthukrishnan, S. } (2005) An improved data stream summary: the count-min sketch and
its applications.   {\it J. Alg.}, {\bf   55}, 58–75.


\bibitem{Dawid:1982}
\textsc{Dawid, A. P.}(1982) The Well-Calibrated Bayesian.
{\it J. Amer. Statist. Assoc.}, {\bf 77}, 605-610.

\bibitem{Dawid:1984}
\textsc{Dawid, A. P.}(1984) Statistical Theory: The Prequential Approach. {\it J. Roy. Stat. Soc. Ser. A}, {\bf 147},  278-292.

\bibitem{Dawid:1992}
\textsc{Dawid, A. P. D.}(1992) Prequential data analysis.  
In:  {\it Current Issues in Statistical Inference: Essays in Honor of D. Basu},
Lecture Notes-Monograph Series, {\bf 17}, 113-126,
Institute of Mathematical Statistics, Beechwood OH.

\bibitem{Dawid:2013}
\textsc{Dawid, A. P.}(2013)  Fundamentals of prequential analysis.
\url{https://www3.stat.sinica.edu.tw/2013frontiers/presentation/29.pdf}

\bibitem{Dawid:2018}
\textsc{Dawid, A. P.}(2018) Bruno de Finetti's Objectivity.
{\it De Finetti Lecture, International Society for Bayes Analysis}.
See: \url{https://bayesian.org/isba-2018-de-finetti-lecture-by-philip-dawid/}

\bibitem{Dawid:Musio:2015}
\textsc{Dawid, A. P. and Musio, M.} Bayesian Model Selection Based on Proper Scoring Rules.
{\it Bayesian Anal.}, {\bf 10} 479 - 499.

\bibitem{Dawid:Vovk:1999}
\textsc{Dawid, A.P. and Vovk, V.}(1999) Prequential probability: principles and
properties. {\it Bernoulli},  {\bf 5}, 125-162.


\bibitem{Dawid:Wang:1993}
\textsc{Dawid, A. P. and Wang, J.}(1993) Fiducial prediction and semi-{B}ayes inference.
{\it Ann. Stat.},  {\bf 21}, 1119-1138.



\bibitem{DeFinetti:1937}
\textsc{De Finetti, B.}(1927) La pr\'{e}vision: ses lois logiques, ses sources subjectives.
{\it Ann. Inst. H. Poincar\'{e}}, {\bf 7}, 1-68.



\bibitem{Dempster:1975}
\textsc{Dempster, A. P.}(1973) Alternatives to least squares in m ultiple regression.  In:  Multivariate Statistical 
Inference. Eds. D. G. Kabe and R. P. Gupta. 25-40.New York, Elsevier.

\bibitem{Dupuis:Robert:2003}
\textsc{Dupuis, J. and Robert, C.}(2003)
Variable selection in qualitative models via an entropic explanatory power.
{\it J. Stat. Planning Inference}, {\bf 111}, 77-94.

\bibitem{Durbin:Koopman:2012}
\textsc{J. Durbin and S. Koopman}(2012) {\it Time Series Analysis by Stat Space Methods.}
Oxford University Press.

\bibitem{Ebrahimi:etal:2010}
\textsc{Ebrahimi, N. Soofi, E., and Soyer, R.}(2010) On the sample information about  parameter and
prediction. {\it Stat. Sci.}, {\bf 25}, 348-367.

\bibitem{Fissler:Presenti:2023}
\textsc{Fissler, T. and Presenti, S.}(2023) Sensitivity mneasures based on scoring functions.
To appear: {\it Euro. J. Operations Res.}



\bibitem{Fong:etal:2022}
\textsc{Fong, E., Holmes, C. and Walker, S.}(2021) Martingale posterior distributions.
\url{arXiv:2103.15671v2}.


\bibitem{Fortini:etal:2000}
\textsc{Fortini, S., Ladelli, L. and Regazzini, E.}(2000) Exchangeability, predictive distributions, and
parametric models.  {\it Sankhya}, {\bf 62} 86-109.

\bibitem{Galton:1907}
\textsc{Galton, F.}(1907) Vox Populi. {\it Nature}, {\bf 75},  450-1.


\bibitem{Ghosal:VanderVaart:2017}
\textsc{Ghosal, S. and van der Vaart, A.}(2017)
{\it Fundamentals of Nonparametric Bayesian Inference.}  Cambridge University Press, Cambridge.

\bibitem{Geisser:1993}
\textsc{Geisser, S.} (1993) {\it Predictive Inference: An Introduction},
CRC Press, Chapman \& Hall.

\bibitem{Gneiting:Raftery:2007}
\textsc{Gneiting,  T.  and Raftery, A.}(2007) Strictly Proper Scoring Rules, Prediction, and Estimation.
{\it J. Amer. Stat. Assoc.}, {\bf 102}, 359-378.

\bibitem{Gneiting:Katzfuss:2014}
\textsc{Gneiting, T. and Katzfuss, M.}(2014) Probabilistic Forecasting.
{\it Ann. Rev.Stat. Appl.},  {\bf 1}, 125-151.

\bibitem{Goutis:Robert:1998}
\textsc{Goutis, C. and Robert, C. }(1998). Model choice in generalised linear models: A bayesian approach
via kullback-leibler projections.  {\it Biometrika}, {\bf 85},29–37.

\bibitem{Gurajala:etal:2021}
\textsc{Gurajala, T. and Choppala, P.  and Meka, J.  and Teal, P.}(2021)
Derivation of the {K}alman filter in a Bayesian filtering perspective.  Proceedings of the
{\it 2nd Inernational Conference on Range Technology (ICORT)}, Chandipur, Balasore, India, p. 1-5.

\bibitem{Hand:Yu:2001}
\textsc{Hand, D. and Yu, K.}(2001) Idiot's Bayes -- not so stupid after all?
{\it Int. Stat. Rev.}, {\bf 69}, 385-398.

\bibitem{Hill:etal:2020}
\textsc{Hill, J. , Linero, A., and Murray, J.}(2020)
{\it Annual Review of Statistics and Its Application}, {\bf 7}, 251-278 .


\bibitem{Huggins:Miller:2023}
\textsc{Huggins, J. and Miller, J.}(2023)
Reproducible model selection using bagged posteriors.  {\it Bayesian Analysis},
{\bf 18}, 79-104.

\bibitem{Jordan:etal:2018}
\textsc{Jordan, A., Kr\"{u}ger, and Lerch, S.}(2018)
{\it Evaluating Probablistic Forecasts with {\sf scoringRules}}
See: \url{arxiv.org/pdf/1709.04743.pdf}

\bibitem{K:M:2007}
\textsc{Kontkanen, P. and Myllymaki, P.}(2007). 
A linear-time algorithm for computing the multinomial stochastic complexity.
{\it  Inform. Process. Lett. }, {\bf 103}, 227–233.


\bibitem{Kontkanen:etal:1997}
\textsc{Kontkanen, P. , Myllym\:{a}ki, P., Silander, T.,Tirri, H.,  and Gru\:{u}wald, P.}(1997)
Comparing Predictive Inference Methods for Discrete Domains.  {\it  PMLR}, {\bf  R1}, 311-318.

\bibitem{Le:Clarke:2016}
\textsc{Le, T. and Clarke, B.}(2016) Using the Bayesian Shtarkov solution for predictions.
{\it Comp. Stat. and Data Anal.}, {\bf 104}, 183-196.

\bibitem{Le:Clarke:2017}
\textsc{Le, T and Clarke, B.}(2017) 
A Bayes Interpretation of Stacking for ${\cal{M}}$-complete and ${\cal{M}}$-open settings.
{\it Bayesian Anal.}, {\bf 12},807-829.


\bibitem{Le:Clarke:2019}
\textsc{Le, T. and C larke, B.}(2019) In praise of partially interpretable predictors.
{\it Stat. Anal. and Data Mining}, {\bf 13}, 113-133.

\bibitem{Le:Clarke:2022}
\textsc{Le, T. and C larke, B.}(2022) Interpreting uninterpretable predictors: kernel methods, Shtarkov solutions,
and random forests.  {\it Stat. Theory and Related Fields}, {\bf 6}, 10-28.

\bibitem{Liang:etal:2008}
\textsc{Liang, F. , Paulo, R., Molina, G., Clyde, M. and Berger, J. O.}(2008)
Mixtures of $g$-priors for Bayesian variable selection.  {\it J. Amer. Stat. Assoc.}, {\bf 103}, 410-423.

\bibitem{Lorbert:etal:2012}
\textsc{Lorbert, A., Blei, D., Schapire, R., and Ramadge, P.}(2012)
A Bayesian boosting model.
\url{https://arxiv.org/pdf/1209.1996.pdf}


\bibitem{Madigan:etal:1996}
\textsc{Madigan, D. , Raftery, A., Volinsky, C., and Hoeting, J.}(1996)
iBayesian m odel averaging. 
{\it Proceedings of the AAAI Workshop on Integrating Multiple Learned Models}, 77--83.

\bibitem{McAlinn:West:2022}
\textsc{McAlinn, K. and West, M.}(2022)
Dynamic Bayesian predictive synthesis in time series forecasting.
\url{https://arxiv.org/pdf/1601.07463.pdf}


\bibitem{Meeker:etal:2017}
\textsc{Meeker, W., Hahn, G., and Escobar, L.}(2017) {\it Statistical Intervals} 2nd Ed.
John Wiley and Sons, Hoboken NJ.

\bibitem{Merhav:Feder:1998}
\textsc{Merhav, N. and Feder, M.}(1998) Universal prediction. {\it Trans. Inform. Theory},
{\bf 44}, 2124-2147.

\bibitem{Muller:etal:2015}
\textsc{M\"{u}ller, P., Quintana, F., Jara, A., Hanson, T.}(2015) {\it Bayesian Nonparametric Data Analysis}
Springer International Publishing, Switzerland.

\bibitem{Muthukrishnan:2005}
\textsc{Muthukrishnan, S.}(2005)  Data Streams: Algorithms and Applications.  
{\it Foundations and Trends in Theoretical Computer Science}, {\bf }, 117-236.

\bibitem{Muthu:2009}
\textsc{Muthukrishnan, S.}(2009)
{\it Data Stream Algorithms}, lecture notes, \url{https://www.cs.mcgill.ca/~denis/notes09.pdf}

\bibitem{OHagan:1995}
\textsc{H'Hagan, A.}(1995) Fractional Bayes factors for Model Comparison.
{\it J. Roy. Stat. Soc.} Ser. B, {\bf 57}, 99-138.

\bibitem{Pena:Walker:2001}
\textsc{Guti\'errez-Pe\~na, E. and Walker, S. G.}(2001)
A Bayesian predictive approach to model selection.
{\it J. Stat. Planning and Inference}, {\bf 93}, 259-276.



\bibitem{Petris:etal:2008}
\textsc{Petris, G. and Petrone, S. and Campagnoli, P.}(2008)
{\it Dynamic Linear Models with R.} Springer, Berlin.

\bibitem{Piironen:Vehtari:2016}
\textsc{Piironen, J. and Vehtari, A.}(2016)
Comparison of Bayesian predictive m ethods for model selection.
{\it Stat. Comp.},{\bf 27}, 711-735.



\bibitem{Raftery:Zheng:2003}
\textsc{Raftery, A. and Zheng, Y.}(2003) Discussion:  Performance of Bayesian Model Averaging.
{\it J. Amer. Statist. Assoc.}, {\bf 98}, 931-938.

\bibitem{Rissanen:1984}
\textsc{Rissanen, J.}(1984) Universal coding,  information, prediction and estimation.
{\it IEEE Trans. Inform.  Theory}, {\bf 30}, 629-636.

\bibitem{Rissanen:1996}
\textsc{Rissanen, J.}(1996) Fisher information and stochastic complexity. {\it Trans. Inform. theory}, {\bf 42}, 
40-47.

\bibitem{Rostek:2010}
\textsc{Rostek, M.}  (2010).    Quantile maximization in decision theory.   
{\it Review of Economic Studies}, {\bf 77},  339-371.

\bibitem{Rufo:etal:2012}
\textsc{Rufo,  M., Mart\'{i}n, J., and P\'{e}rez, C.}(2012)
Log-Linear Pool to Combine Prior Distributions:A Suggestion for a Calibration-Based Approach.
{\it Bayes Analysis}, {\bf 7}, 1-28.

\bibitem{Satuluri:Parthasarathy:2012}
\textsc{Satuluri, V. and Parthasarathy, S.}(2012) Bayesian locality sensitive hashing for fast similarity search.
{\it Proceedings of the VLDB Endowment}, {\bf 5},  430–44.
\url{1https://doi.org/10.14778/2140436.2140440}

\bibitem{Savage:1954}
\textsc{Savage, L. J.}(1954)The Foundations of Statistics. Wiley, NY, 2nd Ed. Dover Publications NY 1972.


\bibitem{Shafer:Vovk:2008}
\textsc{Shafer, G. and Vovk, V.}(2008)  A tutorial on conformal prediction.
{\it J. Mach. Learn. Res.}, {\bf 9}, 371-421.

\bibitem{Shalizi:2007}
\textsc{Shalizi, C.}(2007) Class notes \url{https://www.stat.cmu.edu/~cshalizi/754/notes/lecture-03.pdf}

\bibitem{Shtarkov:1987}
\textsc{Shtarkov, Y.}(1987)Universal sequential coding of single
messages.  {\it Problems in Information Transmission}, {\bf  23}, 3–17.

\bibitem{Silver:2012}
\textsc{Silver, N.}(2012) {\it The Signal and the Noise}, Penguin Press, NY, NY.

\bibitem{Siegel:2016}
\textsc{Siegel, E.}(2016) {\it Predictive Analytics} John Wiley and Sons, Hoboken, NJ.



\bibitem{Skouras:Dawid:1998}
\textsc{Skouras, K. and A. P. Dawid}(1998) On efficient point prediction systems.
{\it J. Roy. Stat. Soc. Ser. B}, {\bf 60}, 765-780.

\bibitem{Skouras:Dawid:1999}
\textsc{Skouras, K. and A. P. Dawid}(1999) On efficient probability forecasting systems.
{\it Biometrika}, {\bf 86}, 765-784.

\bibitem{Smyth:Wolpert:1998}
\textsc{Smyth,  P. and Wolpert, D}(1998)
An evaluation of linearly combining density estimators via stacking. Tech. Rep. 98-25
Information and Computer Science Department, Univ. Irvine.

\bibitem{Wang:etal:2012}
\textsc{Wang, C., Hannig, J. and Iyer, H.}(2012) Fiducial prediction intervals.
{\it J. Stat. Planning and Inference}, {\bf 142}, 1980-1990.

\bibitem{Tian:etal:2022}
\textsc{Tian,  Q. and Nordman, D. and Meeker, W.}(2022)
Methods to compute prediction intervals:  A Review and New Results.
{\it Stat. Sci.}, {\bf 37}, 580-597.

\bibitem{Tipping:2001}
\textsc{Tipping, M.}(2001) Sparse Bayesian Learning and the Relevance Vector Machine.
{\it J. Mach. Learn. Res.}, 211-244.




\bibitem{Erven:etal:2012}
\textsc{van Erven, T., Gr\"{u}nwald, P., de Rooij, S.}(2012) Catching up faster by switching sooner: A
predictive approach to adaptive estimation with an application to the AIC-BIC dilemma.
{\it J. Roy. Stat. Soc. Ser.} B, {\bf 74}, 361-417.

\bibitem{Vehtari:Ojanen:2012}
\textsc{Vehtari, A. and Ojanen, J.}(2012) A Survey of Baysian predictive methods for model assessment, selection,
and comparison.  {\it Statist. Surveys}, {\bf 6}, 142-228.

\bibitem{Dawid_Vovk_Shafer}
\textsc{Vovk H and Shafer G and Dawid APD}(202?) A conversation with Phil Dawid.
{\it Stat. Sci.}


\bibitem{Wallis:2014}
\textsc{Wallis, K.}(2014)  Revisiting Francis Galton's Forecasting Competition.
{\it Stat. Sci.}, {\bf 29}, 420-4.

\bibitem{Wolpert:1992}
\textsc{Wolpert, D. }(1992)
Stacked generalization.  {\it Neural Networks}, {\bf  5},241–259.


\bibitem{Wong:Clarke:2004}
\textsc{Wong, H. and Clarke, B.}(2004)
Improvement over Bayes Prediction in Small Samples in the Presence of Model Uncertainty.
{\it Can. J Stat.}, {\bf 32}, 269-283.

\bibitem{Yang:Zhu:2018}
\textsc{Yang, Z. and Zhu, T.}(2018)
Bayesian selection of misspecified models is overconfident and may cause s
purious posterior probabilities for phylogenetic trees.  {\it Proceedings of the National Academy of Sciences},
{\bf 115}, 1854-1859.

\bibitem{Yao:etal:2018}
\textsc{Yao, Y., Vehtari, A., Simpson, D., and Gelman, A. }(2018). 
Using stacking to average Bayesian predictive distributions (with discussion). 
{\it Bayesian Analysis}, {\bf 13},917–1007.


\bibitem{Yu:Clarke:2011}
\textsc{Yu, C.-Y. and Clarke, B.}(2011) Median loss decision theory.
{\it J. Stat. Planning and Inference}, {\bf 141}, 611-623.

\bibitem{Yu:etal:2013}
\textsc{Yu, Q. , MacEachern, S. and Peruggia, M.}(2013)
Clustered Bayesian Model Averaging.  {\it Bayes Analysis}, {\bf 8}, 883=908.


\bibitem{Zellner:1988}
\textsc{Zellner, A.}(1988) Optimal information processing and Bayes' theorem.
{\it American Statistician}, {\bf 42}, 278-280.






\end{thebibliography}
\end{document}